%% file: main.tex
\documentclass[11pt]{article}

\usepackage{subfiles}
\usepackage{titling}
\usepackage{authblk}
\usepackage{microtype} 
\usepackage{graphicx}
\graphicspath{{figures/}}
\usepackage[numbers]{natbib}
\usepackage{amsmath}
\usepackage{bbm}
\usepackage{mathtools}
\usepackage{algorithm}
\usepackage{algpseudocode}
\newtheorem{definition}{Definition}
\newtheorem{example}{Example}
\newtheorem{assumption}{Assumption}
\newtheorem{theorem}{Theorem}
\newtheorem{corollary}{Corollary}
\usepackage{booktabs}
\usepackage{makecell}
\usepackage{caption}
\usepackage{xcolor}
\usepackage[colorlinks=true,
            linkcolor=blue,
            citecolor=blue,
            urlcolor=blue]{hyperref}

\title{Estimating Effects of Longitudinal Modified Treatment Policies (LMTPs) on Rates of Change in Health Outcomes}
\author[1]{Anja Shahu%
    \thanks{Corresponding author: \texttt{as6798@cumc.columbia.edu}}%
    \thanks{This is the author-prepared version of an article published in \emph{Statistics in Medicine}. The published article is available at \url{http://dx.doi.org/10.1002/sim.70604}.}%
}
\author[1]{Weijie Xia}
\author[1]{Ying Wei}
\author[1]{Daniel Malinsky}
\affil[1]{Department of Biostatistics, Mailman School of Public Health, \protect\\ Columbia University, New York, NY, USA}

\begin{document}

\maketitle

\begin{abstract}
    Longitudinal data often contains outcomes measured at multiple visits, and scientific interest may lie in quantifying the effect of an intervention on an outcome's rate of change. For example, one may wish to study the progression (or trajectory) of a disease over time under different hypothetical interventions. We extend the longitudinal modified treatment policy (LMTP) methodology to estimate effects of complex, exposure-dependent interventions on rates of change in an outcome over time. We exploit the theoretical properties of a nonparametric efficient influence function (EIF)-based estimator to introduce a novel inference framework that can be used to construct simultaneous confidence intervals for a variety of causal effects of interest and to formally test relevant global and local hypotheses about rates of change. We demonstrate the utility of our framework in investigating whether a longitudinal shift intervention affects an outcome's counterfactual trajectory, as compared with no intervention. We present results from a simulation study to illustrate the performance of our inference framework in a longitudinal setting with time-varying confounding and a continuous exposure. We also apply our inference framework to the Columbia Brain Health DataBank (CBDB) to examine the effect of shifting blood pressure on the progression of dementia.
\end{abstract}

\newpage

\section{Introduction}
\label{sec:intro}

Longitudinal studies often measure outcomes repeatedly over time, and scientific interest frequently centers on rates of change in these outcomes. For example, investigators may wish to evaluate how hypothetical interventions affect outcome trajectories or progression over time (e.g., effects on rate of cognitive decline or progression of emphysema). However, such questions lack a clear and flexible causal formulation in the existing literature. Existing approaches have typically targeted simpler estimands, such as average effects on outcomes at the end of the study, or have relied on parametric notions of rate of change. For example, in the marginal structural model (MSM) literature, effects on rates of change are tied to exposure-by-time interaction coefficients in parametric regression models \citep{petersen2007history, malinsky2025marginal}. These parameters have no independent meaning or clear interpretation in the absence (or misspecification) of the assumed parametric model \citep{vansteelandt2022assumption}. Moreover, these approaches are generally limited to the evaluation of static interventions. In simpler pre-post settings, investigators have also considered change score approaches to study effects on changes in outcomes. However, these approaches have been criticized for not representing well-defined causal estimands in observational studies \citep{glymour2005baseline, shahar2012causal, tennant2022analyses}. A related area of work considers curve effects, but these approaches are generally interested in curves as functions of the exposure or intervention, rather than time, and thus do not capture how outcome trajectories evolve over time \citep{kennedy2017non, kennedy2019nonparametric, bonvini2022fast}. For example, \citet{kennedy2019nonparametric} formalizes an incremental effect curve to summarize effects under a continuum of interventions that shift propensity score values, characterizing how the outcome varies with treatment intensity. In this work, we build on the flexible, nonparametric longitudinal modified treatment policy (LMTP) framework to study the effects of complex, exposure-dependent interventions on rates of change in outcomes over time.

LMTPs can be viewed as a generalization of dynamic treatment regimes that allow the intervention to depend on the natural value of exposure, which is the value that a time-varying exposure would take at some time if the intervention was discontinued right before that time. This type of intervention has been formalized in several prior papers \citep{diaz2023nonparametric, robins2004effects, munoz2012population, richardson2013single, young2014identification, haneuse2013estimation}, but our work builds primarily on the methodology developed by \citet{diaz2023nonparametric}. Some examples of LMTPs include shift interventions (e.g., shifting exposure by a constant), threshold interventions (e.g., setting exposure to a value when it falls below a specified threshold), or certain stochastic interventions (e.g., drawing exposure from a shifted distribution to represent a population-level shift). Since the LMTP methodology provides an unified framework to evaluate a broad class of interventions, static and dynamic treatment regimes can be studied as special cases, even though they do not explicitly depend on the natural value of exposure. With its ability to accommodate such a wide range of interventions, the LMTP methodology constitutes an ideal foundation to build upon.

The LMTP framework also has several other desirable qualities, particularly for longitudinal data. First, an LMTP approach allows researchers to evaluate policy-relevant interventions and encourages researchers to formulate estimands that correspond to interesting, complex research questions. Second, the LMTP methodology properly accounts for time-varying confounding, a complication that is often encountered in longitudinal data. Third, the LMTP framework affords flexibility in the type of exposures being considered, allowing for continuous and categorical exposures as well as binary ones. Fourth, the LMTP methodology proposes efficient influence function (EIF)-based estimators that have desirable theoretical guarantees and allow for flexible, data-adaptive estimation. Finally, an LMTP approach facilitates a focus on estimands that are more likely to satisfy the positivity assumption, which is key to reliable causal inference. 

Recent work on LMTP effects has focused either on outcomes measured at the end of a study or survival outcomes \citep{diaz2023nonparametric, diaz2024causal}. In this paper, we extend the LMTP methodology to a new and practically important class of causal targets: effects on rates of change in outcomes. We formalize these causal targets independently of parametric modeling assumptions, that is, as nonparametric functionals of the data-generating distribution under LMTP interventions. By expressing these targets as linear contrasts of outcome trajectories, we develop a principled inference framework to conduct global and simultaneous inference for effects on rates of change. 

The paper is organized as follows: Section~\ref{sec:lmtps} provides an introduction to LMTPs; Section~\ref{sec:causal} formalizes causal effects on rates of change; Section~\ref{sec:inf} proposes an inference framework and outlines its usefulness in investigating whether the rate of change in an outcome changes when an LMTP is implemented; Section~\ref{sec:sim} investigates the performance of the framework in a longitudinal setting using simulated data; Section~\ref{sec:app} presents an illustrative application of the framework using the Columbia Brain Health DataBank (CBDB); and finally, Section~\ref{sec:discussion} discusses limitations of the framework and proposes some future areas of work.  

\section{LMTPs}
\label{sec:lmtps}

In this section, we provide an introduction to LMTPs and establish relevant notation. Our setup is similar to \citet{diaz2023nonparametric} but extended to accommodate a continuous outcome that is measured repeatedly over time. Consider a sample of independent and identically distributed (iid) observations $Z_1, \ldots, Z_n \sim P$, where distribution $P$ is contained in a nonparametric statistical model. Let $Z = (L_1, A_1, Y_1, \ldots, L_\tau, A_\tau, Y_\tau)$. For discrete time $t \in \{1, \ldots, \tau\}$, $L_t$ is a vector of time-varying covariates, $A_t$ is a time-varying exposure, and $Y_t$ is a continuous time-varying outcome. Here, the intervals between successive time points are assumed to be equally spaced (e.g., one year apart) and regular for each individual. For a random variable $X$, denote its history and future as $\overline{X}_t = (X_1, \ldots, X_t)$ and $\underline{X}_t = (X_t, \ldots, X_\tau)$, respectively. We use $\overline{X}$ to denote the complete history $(X_1, \ldots, X_\tau)$. Denote the history of all random variables until just before $A_t$ as $H_t = (\overline{A}_{t-1}, \overline{L}_t, \overline{Y}_{t-1})$. 

The causal model is formalized via a non-parametric structural equation model. This means data is generated based on deterministic functions $f_{L_t}$, $f_{A_t}$, and $f_{Y_t}$, such that
\begin{align*}
  &L_t = f_{L_t}(A_{t-1}, Y_{t-1}, H_{t-1}, U_{L,t}) \\
  &A_t = f_{A_t}(H_t, U_{A,t}) \\
  &Y_t = f_{Y_t}(A_t, H_t, U_{Y,t}),
\end{align*}
where $U = (U_{L,t}, U_{A,t}, U_{Y,t}: t \in \{1, \dots, \tau \})$ is the set of exogenous variables. This implies a strict time-ordering at each time $t$ (i.e., $L_t \rightarrow A_t \rightarrow Y_t$).

We define an intervention as replacing $A_t$ in the structural model with $A_t^d$. Intervening on all exposures up until time $t-1$ is denoted by $\overline{A}^d_{t-1} = (A_1^d, \ldots, A_{t-1}^d)$ and induces the following counterfactual variables
\begin{align*}
  &L_t(\overline{A}^d_{t-1}) = f_{L_t}(A^d_{t-1}, Y_{t-1}(\overline{A}^d_{t-1}), H_{t-1}(\overline{A}^d_{t-2}), U_{L,t}) \\
  &A_t(\overline{A}^d_{t-1}) = f_{A_t}(H_t(\overline{A}^d_{t-1}), U_{A,t}) \\
  &Y_{t-1}(\overline{A}^d_{t-1}) = f_{Y_{t-1}}(A^d_{t-1}, H_{t-1}(\overline{A}^d_{t-2}), U_{Y,t-1}),
\end{align*}
where $H_t(\overline{A}^d_{t-1}) = (\overline{A}^d_{t-1}, \overline{L}_t(\overline{A}^d_{t-1}), \overline{Y}_{t-1}(\overline{A}^d_{t-1}))$ is the counterfactual history. We refer to $A_t(\overline{A}^d_{t-1})$ as the natural value of exposure; that is, the value that the time-varying exposure would take at time $t$ if the intervention was discontinued right before time $t$. The counterfactual outcome $Y_{t-1}(\overline{A}^d_{t-1})$ can be interpreted similarly. 

\begin{definition}[LMTP]
  \label{def:lmtp}
  The intervention $A_t^d$ is called a LMTP if it is defined as $A_t^d = d(A_t(\overline{A}^d_{t-1}), H_t(\overline{A}^d_{t-1}))$ for a user-given function $d$.
\end{definition}

An LMTP is defined by function $d$ that takes both the natural value of exposure and counterfactual history as inputs. One example of a common LMTP is an additive or multiplicative shift intervention, in which the natural value of exposure is modified by some constant $\delta$ at each time point. Rather than shifting all individuals, we may consider shifting individuals such as in Example~\ref{ex:shift_lmtp} so that the shifted values fall inside the range of the empirical values, ensuring the positivity assumption holds by design.
\begin{example}[Additive Shift LMTP]
  \label{ex:shift_lmtp}
  Let $A_t$ be air pollution concentration at time $t$. Suppose $u_t$ exists such that $P(A_t > u_t \mid H_t = h_t) = 1$ for all $t \in \{1, \ldots, \tau\}$. For constant $\delta$, define
  \begin{align*}
    d(a_t, h_t)= \begin{cases} a_t - \delta, & \text { if } a_t - \delta \geq u_t(h_t) \\ a_t, & \text { if } a_t - \delta < u_t(h_t) \end{cases}
  \end{align*}
  which can be interpreted as reducing the air pollution concentration input by $\delta$ only if the resulting shifted value falls above the minimum of the empirical values.
\end{example}

In this paper, we will often refer to this kind of shift intervention as our LMTP of interest; however, the theory developed can apply to other LMTPs that satisfy Definition~\ref{def:lmtp} \citep{diaz2023nonparametric, hoffman2024studying}. Specifically, it applies to LMTPs that are piecewise smooth invertible. This means that either $A_t$ must be discrete for all $t$ or if $A_t$ is continuous, $d(a_t,h_t)$ must be piecewise smooth invertible with respect to $a_t$. This assumption of piecewise smooth invertibility is required to construct EIF-based estimators with nice theoretical properties. A threshold intervention, for example, is not pathwise differentiable and would thus violate this assumption and fall outside the scope of the present theory.

\section{Causal Effects on Rates of Change}
\label{sec:causal}

For a particular time $t$, consider the following causal estimand
\begin{align*}
  \theta_t = E[Y_t(\overline{A}^d_t)],
\end{align*}
which is the expected counterfactual outcome at time $t$ under an intervention defined by function $d$. We use $\theta'_t$ and $\theta''_t$ to distinguish estimands corresponding to two different interventions $d'$ and $d''$. These functions can be any type of intervention, including no intervention.

\begin{definition}[Counterfactual outcome trajectory]
  The vector of causal estimands across time $t \in \{1, \ldots, \tau \}$, denoted as $\overline{\theta} = (\theta_1, \ldots, \theta_\tau)$, is called the counterfactual outcome trajectory. If $d$ is no intervention, we call this vector the natural outcome trajectory to emphasize that it involves no counterfactuals.
\end{definition}

For some intervention $d$, $\theta_{t} - \theta_{1}$ represents the change in the expected counterfactual outcome in the interval from baseline to time $t$. Thus, we can define a relevant causal effect on the outcome's rate of change, comparing two interventions $d'$ and $d''$, as 
\begin{align*}
  \Delta_t = \theta''_{t} - \theta''_{1} - (\theta'_{t} - \theta'_{1}).
\end{align*}
In other words, this is the difference in the change in the outcome, on average, in the interval from baseline to time $t$, comparing $d''$ to $d'$. Here, we use the term ``rate of change'' in the discrete-time sense (over a fixed interval length), rather than to represent a rate in the sense of change per unit time. Since individuals are measured at a common set of times, dividing by the length of the interval would just rescale $\Delta_t$ and would not change its causal interpretation, making normalization unnecessary in the discrete-time setting. 

\begin{definition}[Causal effects]
  The vector of causal effects for all combinations of time points $t \in \{2, \ldots, \tau\}$ with baseline time $t = 1$ is $\Delta = (\Delta_2, \ldots, \Delta_\tau)$.
\end{definition}  

The vector of causal effects $\Delta$ compares how the counterfactual outcome trajectory evolves over time under $d'$ versus $d''$. For simplicity, we let $d'$ be no intervention and $d''$ be a LMTP intervention for the rest of this paper, so our interest is how an LMTP changes the counterfactual outcome trajectory, as compared with no intervention. Alternative definitions of $\Delta_t$ could also be considered. For example, $\Delta_t$ could be defined to compare adjacent time points by swapping out $\theta'_{1}$ and $\theta''_{1}$ for $\theta'_{t-1}$ and $\theta''_{t-1}$, respectively. The framework we propose in the next section will allow for flexibility in the way $\Delta_t$ is defined. 

Theory for identification and nonparametric estimation of the vector of causal effects $\Delta$ follows straightforwardly from results established for $\theta_t$ in \citet{diaz2023nonparametric}. The following assumptions are sufficient for identification of $\theta_t$. 
\begin{assumption}[Positivity] \label{assump:pos}
    If $(a_s,h_s) \in \text{supp}\{A_s, H_s\}$, then $(d(a_s, h_s), h_s)$ \\ $\in \text{supp}\{A_s, H_s\}$ for $s \in \{1, \ldots, t\}$. In words, if there is an individual at time $s$ with history $h_s$ and exposure $a_s$, then it is also possible to find an individual with history $h_s$ and exposure $d(a_s,h_s)$.
\end{assumption} 
\begin{assumption}[Strong sequential randomization] \label{assump:strong}
    $U_{A,s} \perp\!\!\!\perp (\underline{U}_{Y,s},$ \\ $\underline{U}_{L,s+1},\underline{U}_{A,s+1}) \mid H_s$ for all $s \in \{1, \ldots, t\}$. This is satisfied at time $s$ if $H_s$ includes all common causes of the intervention variable $A_s$ and $(Y_s, L_{s+1}, A_{s+1},$ $\ldots,Y_{t-1}, L_t, A_t, Y_t)$.
\end{assumption}

A g-computation identification expression is given in terms of sequential regressions. For $s = t, \ldots, 1$, recursively define the sequential regressions
\begin{align*}
  m_{s, t}(a_s, h_s) = E\left[ m_{s+1, t}(A_{s+1}^d, H_{s+1}) \mid A_s = a_s, H_s = h_s\right],
\end{align*}
starting with $m_{t+1, t}(A_{t+1}^d, H_{t+1}) = Y_t$. Then, under Assumptions~\ref{assump:pos} and~\ref{assump:strong}, $\theta_t$ is identified as
\begin{align*}
    \theta_t = E \left[ m_{1,t}(A_1^d, L_1) \right].
\end{align*}
For certain interventions that are a subclass of LMTPs, such as static or dynamic treatment regimes, a weaker version of Assumption~\ref{assump:strong} would be sufficient for identification. This assumption is referred to as ``standard sequential randomization'' in \citet{diaz2023nonparametric}. Formally, this assumption is that $U_{A,s} \perp\!\!\!\perp (\underline{U}_{Y,s}, \underline{U}_{L,s+1}) \mid H_s$ for all $s \in \{1, \ldots, t\}$, which is satisfied at time $s$ if $H_s$ includes all common causes of the $A_s$ and $(Y_s, L_{s+1}, \ldots, Y_{t-1}, L_t, Y_t)$. See \cite{diaz2023nonparametric} for additional details. 

A simple estimator for $\theta_t$ motivated by the identification result would be a sequential regression g-computation estimator, which recursively regresses estimates of $m_{s+1, t}(A_{s+1}^d, H_{s+1})$ on $(A_s, H_s)$ from $s = t, \ldots, 1$ to obtain the corresponding estimates of $m_{s,t}$. However, this estimator does not attain the desirable theoretical guarantees of EIF-based estimators, and we therefore do not pursue it any further. Instead, we consider the two EIF-based estimators for $\theta_t$ that were proposed in \citet{diaz2023nonparametric}: the sequential doubly robust (SDR) estimator and the targeted minimum loss-based estimator (TMLE). Under appropriate conditions, both estimators are $n^{1/2}$-consistent and asymptotically normal. The framework we propose in the next section builds directly on these properties. These estimators are also doubly robust, which refers to their ability to remain consistent even if some nuisance components are inconsistently estimated. Nuisance functions in both the SDR estimator and TMLE are estimated using flexible data-adaptive techniques, such as Super Learner, that reduce the risk of model misspecification while still preserving proper convergence rates \citep{van2007super}. For the purpose of our simulation study and illustrative application, we use a simple set of candidate learners in the Super Learner for computational feasibility (i.e., generalized linear model [GLM], intercept-only model, multivariate adaptive regression splines [MARS], and generalized additive model [GAM]). Our implementation for estimating $\Delta$ follows straightforwardly from the functions available to estimate $\theta_t$ in the \texttt{lmtp} R package \citep{williams2023lmtp}. In particular, we can estimate $\Delta$ by running the estimation algorithm for $\theta_1, \dots, \theta_{\tau}$ (i.e., a loop of $\tau$ estimation runs) for both $d'$ and $d''$. 

\section{Inference Framework}
\label{sec:inf}

So far, we have been able to rely on the existing LMTP infrastructure to estimate the vector of causal effects $\Delta$. However, additional theory and implementation is needed to conduct relevant inference for this vector. In this section, we propose a comprehensive inference framework that draws inspiration from \citet{hothorn2008simultaneous}. We outline global and simultaneous inference that can be conducted with this framework that is relevant for investigating whether the rate of change in an outcome differs under an LMTP intervention compared with no intervention. Specifically, we use this framework to formally test for both global and local differences in the rate of change in the counterfactual outcome trajectory $\overline{\theta}''$ versus the natural outcome trajectory $\overline{\theta}'$ and to construct simultaneous confidence intervals for the vector of causal effects $\Delta$, which quantifies the magnitude of difference. 

\subsection{Building Intuition}

\begin{figure}[!tbp]
  \centering
  \includegraphics[width=\textwidth]{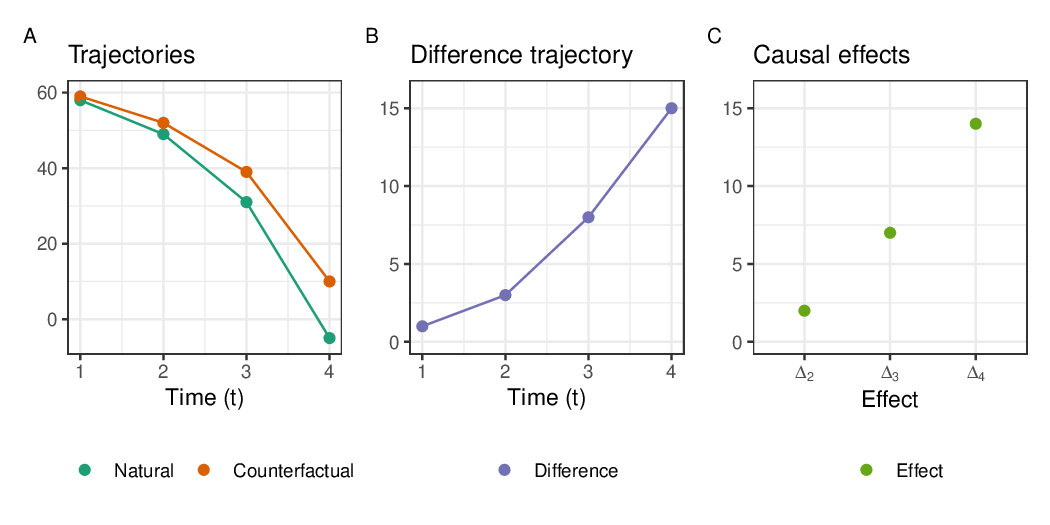}
  \caption{Synthetic data example comparing a potential natural outcome trajectory $\overline{\theta}'$ and counterfactual outcome trajectory $\overline{\theta}''$, in which a difference in the rates of change exists. Panel A visualizes the two decreasing trajectories. Panel B visualizes the difference trajectory obtained by subtracting $\overline{\theta}'$ from $\overline{\theta}''$. Panel C visualizes the vector of causal effects $\Delta$ obtained by subtracting each point of the difference trajectory by the baseline point.}
  \label{fig:inf_visual}
\end{figure}

Figure~\ref{fig:inf_visual} provides intuition for how we construct our inference framework. Using a synthetic data example, we plot the corresponding natural outcome trajectory $\overline{\theta}'$ and counterfactual outcome trajectory $\overline{\theta}''$ for comparison in Panel A. Our goal is to investigate whether these two trajectories differ in terms of their rate of change. In this example, there is a clear difference, as the two trajectories are non-parallel, with $\overline{\theta}'$ declining more rapidly than $\overline{\theta}''$. First notice that when we take the difference between the two trajectories, we obtain the difference trajectory, as shown in Panel B, and find that this trajectory is non-constant over time. Subtracting each post-baseline point of the difference trajectory from the baseline point, we obtain the vector of causal effects $\Delta$, as shown in Panel C, and find that the components of $\Delta$ are non-zero. Therefore, one simple approach for globally testing whether two estimated trajectories differ significantly in terms of their rates of change would be to test whether at least one component of the estimated $\Delta$ vector is non-zero. If this global test is significant, we may proceed with local tests, individually testing which components of the estimated $\Delta$ are non-zero and providing these estimates and their simultaneous confidence intervals to quantify the magnitude of difference. 

Since our target for inference, the vector of causal effects $\Delta$, is comprised of linear combinations of estimands, we formulate our inference framework in terms of linear functions of estimands using matrix algebra. We use this more general formulation because it affords the investigator flexibility in the way $\Delta$ is defined. Since many useful causal quantities are in fact just linear combinations of estimands, this makes our framework applicable for conducting inference for other causal quantities that may be of interest but are not the focus of this paper.

\subsection{Set Up}

Consider stacking the natural outcome trajectory $\overline{\theta}'$ and the counterfactual outcome trajectory $\overline{\theta}''$ into a single vector $\theta \in \mathbbm{R}^{2\tau \times 1}$ such that
\begin{align*}
  \theta = 
  \begin{pmatrix} 
    \theta'_1 & \cdots & \theta'_\tau & \theta''_1 & \cdots & \theta''_\tau 
  \end{pmatrix}^T.
\end{align*}

Suppose we focus on constructing an estimator for $\theta$. For the purpose of illustrating this framework, we choose to use the SDR estimator instead of the TMLE because the SDR estimator is more robust to model misspecification in the sense that it is $2^\tau$ doubly robust, while the TMLE is only $\tau + 1$ doubly robust \citep{diaz2023nonparametric, hoffman2024studying}. The former type of double robustness guarantees consistency of $\theta_t$ if either the outcome regression or treatment mechanism is consistently estimated for each time $s \in \{1, \ldots, t\}$, while the latter type guarantees consistency if for some time $k$, all outcome regressions for $s > k$ and all treatment mechanisms for $s \leq k$ are consistently estimated. In contrast to the TMLE, which enforces parameter space constraints by construction, the default implementation of the SDR estimator can yield estimates that fall outside the bounds of the parameter space. However, this issue can be addressed without compromising the asymptotic properties of the SDR estimator by projecting the pseudo-outcome regressions onto the space of functions that respect the parameter space via isotonic regression, as in \citet{susmann2025computationally}. Alternatively, one could consider using the TMLE instead. Even though the estimation algorithm for the TMLE differs from the SDR estimator, its construction based on the EIF allows a similar weak convergence result to be established so that the theory developed for inference still holds.

We will now provide a brief overview of how the SDR estimator for $\theta_t$ from \citet{diaz2023nonparametric} is constructed. Recall that the outcome regression $m_{s,t}$ was defined recursively in the previous section. It is helpful to also define the density ratio for $s \in \{1, \ldots, t\}$
\begin{align*}
      r_{s,t}(a_s, h_s) = \frac{g_s^d(a_s \mid h_s)}{g_s(a_s \mid h_s)},
\end{align*}
where $g_s(a_s \mid h_s)$ and $g_s^d(a_s \mid h_s)$ are the densities of $A_s$ and $A_s^d$, respectively, conditional on $H_s = h_s$. Additional details on how the post-intervention density $g_s^d(a_s \mid h_s)$ is defined using the piecewise smooth invertibility assumption can be found in \citet{diaz2023nonparametric}.

We also define the data transformation for $s \in \{1, \ldots, t\}$ 
\begin{align*}
    \phi_{s,t}(z; \underline{\eta}_{s,t}) &= \sum_{p=s}^t \left( \prod_{k=s}^p r_{k,t}(a_k, h_k) \right) \left\{m_{p+1, t}(a_{p+1}^d, h_{p+1}) - m_{p,t}(a_p,h_p)\right\} \\
    &\quad + m_{s,t}(a_s^d, h_s),
\end{align*}
which depends on the vector of nuisance parameters $\underline{\eta}_{s,t} = (r_{s,t}, m_{s,t}, \ldots,$ $ r_{t,t}, m_{t,t} )$. Notice that $\phi_{s,t}$ has a form that resembles an EIF. In fact, it turns out that the EIF for $\theta_t$ in the nonparametric model is given by $\phi_{1,t}(Z; \eta_t) - \theta_t$, where $\eta_t$ is used to denote the entire vector of nuisance parameters $\left(r_{1,t}, m_{1,t}, \ldots, r_{t,t}, m_{t,t} \right)$. Thus, an estimator for $\theta_t$ could be constructed by averaging $\phi_{1,t}(Z; \widehat{\eta}_t)$ across the sample. 

In \citet{diaz2023nonparametric}, it was shown that $\phi_{s+1, t}$ is a multiply robust unbiased transformation for $m_{s, t}$ in the sense that 
\begin{align*}
    E \left[ \phi_{s+1,t}(Z; \underline{\eta}_{s+1, t}') \Bigm| A_s = a_s, H_s = h_s \right] = m_{s,t}(a_s, h_s)
\end{align*}
if $\underline{\eta}_{s+1,t}'$ is such that either $m_{k,t}' = m_{k, t}$ or $r_{k, t}' = r_{k, t}$ for each $k \geq s+1$ (i.e., if either the outcome regressions or the exposure density ratios are consistently estimated at time points after $s$). In other words, the conditional expectation of $\phi_{s+1,t}$ equals $m_{s,t}$ even if some nuisance parameters are inconsistently estimated. This property motivates constructing the SDR estimator for $\theta_t$ by recursively regressing estimates of $\phi_{s+1,t}(Z; \underline{\eta}_{s+1, t})$ on $(A_s, H_s)$ from $s = t, \ldots, 1$ to obtain the corresponding estimates of $m_{s,t}$. Such an algorithm is outlined in Algorithm~\ref{alg:sdr}. The SDR estimator for $\theta_t$ is defined as
\begin{align*}
  \widehat{\theta}_t = \frac{1}{n} \sum_{i=1}^n \phi_{1,t}(Z_i, \widehat{\eta}_{i,t}),
\end{align*}
where $i$ indexes individuals in the sample. In the special case where $d$ is no intervention, $\phi_{1,t}(Z_i, \widehat{\eta}_{i,t}) = Y_{i,t}$. 

For our purposes, we use the SDR estimator from \citet{diaz2023nonparametric} without any modification to its construction. The only feature specific to our setting is that we evaluate this estimator under two interventions $d'$ and $d''$ and across time $t \in \{1, \ldots, \tau\}$, and we stack the resulting estimates into a single vector. Thus, the SDR estimator for $\theta$ is given by $\widehat{\theta}_n \in \mathbbm{R}^{2\tau \times 1}$ such that 
\begin{align*}
  \widehat{\theta}_n = 
  \begin{pmatrix}
    \frac{1}{n} \sum_{i=1}^n \widehat{\phi}'_{i,1} & \cdots & \frac{1}{n} \sum_{i=1}^n \widehat{\phi}'_{i,\tau} & \frac{1}{n} \sum_{i=1}^n \widehat{\phi}''_{i,1} & \cdots & \frac{1}{n} \sum_{i=1}^n \widehat{\phi}''_{i,\tau}
  \end{pmatrix}^T,
\end{align*}
where $\widehat{\phi}_{i,t}$ is used as shorthand for $\phi_{1,t}(Z_i, \widehat{\eta}_{i,t})$ for notational simplicity. We also similarly use $\phi_t$ as shorthand for $\phi_{1,t}(Z, \eta_t)$ later.

\begin{algorithm}[H]
\caption{SDR estimator for $\theta_t$}
\label{alg:sdr}
\begin{algorithmic}[1]
\State Initialize pseudo-outcome $\phi_{t+1,t}(Z;\underline{\widehat{\eta}}_{t+1,t}) = Y_t$
\For{$s = t, \ldots, 1$}
    \State Regress $\phi_{s+1,t}(Z;\underline{\widehat{\eta}}_{s+1,t})$ on $(A_s, H_s)$
    \State Compute predictions from the fitted regression at $A_s^d = d(A_s,H_s)$
    \Statex \hspace{2.5em} and $A_s$, that is, $\widehat{m}_{s,t}(A_s^d, H_s)$ and $\widehat{m}_{s,t}(A_s, H_s)$
    \State Compute $\widehat{r}_{s,t}(A_s, H_s)$ via classification approach in \citet{diaz2023nonparametric}
    \State Calculate pseudo-outcome $\phi_{s,t}(Z;\underline{\widehat{\eta}}_{s,t})$
\EndFor
\State Obtain $\widehat{\theta}_t$ by averaging $\phi_{1,t}(Z;\widehat{\eta}_t)$ across the sample
\end{algorithmic}
\end{algorithm}

The SDR estimator $\widehat{\theta}_n$ satisfies a weak convergence result that follows directly from Theorem 4 in \citet{diaz2023nonparametric}, which states that under proper convergence rates for the nuisance parameters, $\widehat{\theta}_t$ weakly converges to a normal distribution with variance defined by the EIF. This convergence result requires a Donsker-type assumption that imposes complexity restrictions or the use of cross-fitting for nuisance estimation \citep{van2000asymptotic, chernozhukov2018double}. To avoid Donsker classes altogether, we use cross-fitting. 

\begin{theorem}[Weak convergence of SDR estimator for $\theta$]
  \label{thm:weak_converg_theta}
  Assume conditions of Theorem 2 and 4 from \citet{diaz2023nonparametric} hold. Then, 
  \begin{align*}
    \sqrt{n}(\widehat{\theta}_n - \theta) \overset{d}{\longrightarrow} N_{2\tau}(0, \Sigma),
  \end{align*}
  where $0 \in \mathbbm{R}^{2\tau \times 1}$ is a vector of zeros and $\Sigma \in \mathbbm{R}^{2\tau \times 2\tau}$ is a symmetric covariance matrix defined as
  \begin{align*}
    \pmb{\Sigma} = 
    \begin{pmatrix}
      \sigma^2_{\phi'_1} &  &  &  \\
      \sigma_{\phi'_2, \phi'_1} & \sigma^2_{\phi'_2} &  &  \\
      \vdots  & \vdots  & \ddots &  \\
      \sigma_{\phi'_\tau, \phi'_1} &  \sigma_{\phi'_\tau, \phi'_2} & \cdots & \sigma^2_{\phi'_\tau} \\
      \sigma_{\phi''_1, \phi'_1} & \sigma_{\phi''_1, \phi'_2} & \cdots & \sigma_{\phi''_1, \phi'_\tau} & \sigma^2_{\phi''_1} & \\
      \sigma_{\phi''_2, \phi'_1} & \sigma_{\phi''_2, \phi'_2} & \cdots & \sigma_{\phi''_2, \phi'_\tau} & \sigma_{\phi''_2, \phi''_1} & \sigma^2_{\phi''_2} & \\
      \vdots & \vdots & \vdots & \vdots & \vdots & \vdots & \ddots & \\
      \sigma_{\phi''_\tau, \phi'_1} & \sigma_{\phi''_\tau, \phi'_2} & \cdots & \sigma_{\phi''_\tau, \phi'_\tau} & \sigma_{\phi''_\tau, \phi''_1} & \sigma_{\phi''_\tau, \phi''_2} & \cdots & \sigma^2_{\phi''_\tau}
    \end{pmatrix},
  \end{align*}
  where for some random variables $X$ and $W$, $\sigma^2_X = \text{Var}(X)$ and $\sigma_{X,W} = \text{Cov}(X,W)$.
\end{theorem}

Suppose $S_n \in \mathbbm{R}^{2\tau \times 2\tau}$ is a consistent estimator of $\frac{1}{n} \Sigma$, which means that 
\begin{align}
  n S_n \overset{p}{\longrightarrow} \Sigma.
  \label{eqn:S_converg}
\end{align}
For example, $\Sigma$ may be estimated using the empirical covariance matrix of $\widehat{\phi} \in \mathbbm{R}^{n \times 2\tau}$, which is defined as
\begin{align*}
  \widehat{\phi} = 
  \begin{pmatrix}
    \widehat{\phi}'_{1,1} & \cdots & \widehat{\phi}'_{1,\tau} & \widehat{\phi}''_{1,1} & \cdots & \widehat{\phi}''_{1,\tau} \\
    \vdots & \vdots & \vdots & \vdots & \vdots & \vdots \\
    \widehat{\phi}'_{n,1} & \cdots & \widehat{\phi}'_{n,\tau} & \widehat{\phi}''_{n,1} & \cdots & \widehat{\phi}''_{n,\tau}
  \end{pmatrix}.
\end{align*}
Then, by Slutsky's Theorem, the approximate distribution of $\widehat{\theta}_n$ is a multivariate normal distribution with mean $\theta$ and covariance matrix $S_n$, or 
\begin{align}
  \widehat{\theta}_n \overset{app}{\sim} N_{2\tau}(\theta, S_n). 
  \label{eqn:mvn_theta_hat}
\end{align}

Now consider a linear function of $\theta$, denoted by
\begin{align*}
  \nu = K \theta \in \mathbbm{R}^{k \times 1},
\end{align*}
for some user-given constant matrix $K \in \mathbbm{R}^{k \times 2\tau}$.

\begin{example}[Choosing $K$ to yield $\Delta$]
  \label{ex:K_roc}
  Let 
  \begin{align*}
    K = 
    \begin{pmatrix} 
      1 & -1 & 0 & 0 & \cdots & 0 & 0 & -1 & 1 & 0 & 0 & \cdots & 0 & 0 \\ 
      1 & 0 & -1 & 0 & \cdots & 0 & 0 & -1 & 0 & 1 & 0 & \cdots & 0 & 0 \\
      \vdots & \vdots & \vdots & \vdots & \vdots & \vdots & \vdots & \vdots & \vdots & \vdots & \vdots & \vdots & \vdots & \vdots \\
      1 & 0 & 0 & 0 & \cdots & 0 & -1 & -1 & 0 & 0 & 0 & \cdots & 0 & 1
    \end{pmatrix}. 
  \end{align*} 
  Then, 
  \begin{align*}
    \nu 
    = \begin{pmatrix} 
      \theta'_1 - \theta'_2 - \theta''_1 + \theta''_2 \\
      \theta'_1 - \theta'_3 - \theta''_1 + \theta''_3 \\
      \vdots \\
      \theta'_1 - \theta'_\tau - \theta''_1 + \theta''_\tau
    \end{pmatrix} 
    = \begin{pmatrix} 
      \theta''_2 - \theta''_1 - (\theta'_2 - \theta'_1) \\
      \theta''_3 - \theta''_1 - (\theta'_3 - \theta'_1) \\
      \vdots \\
      \theta''_\tau - \theta''_1 - (\theta'_\tau - \theta'_1)
    \end{pmatrix}
    = \begin{pmatrix}
      \Delta_2 \\
      \Delta_3 \\
      \vdots \\
      \Delta_\tau
    \end{pmatrix}
  \end{align*}
  which is the vector of causal effects $\Delta$.
\end{example}

\begin{example}[Choosing $K$ to yield alternate version of $\Delta$]
  \label{ex:K_roc_alt}
  Let 
  \begin{align*}
    K = 
    \begin{pmatrix} 
      1 & -1 & 0 & 0 & \cdots & 0 & 0 & -1 & 1 & 0 & 0 & \cdots & 0 & 0 \\ 
      0 & 1 & -1 & 0 & \cdots & 0 & 0 & 0 & -1 & 1 & 0 & \cdots & 0 & 0 \\
      \vdots & \vdots & \vdots & \vdots & \vdots & \vdots & \vdots & \vdots & \vdots & \vdots & \vdots & \vdots & \vdots & \vdots \\
      0 & 0 & 0 & 0 & \cdots & 1 & -1 & 0 & 0 & 0 & 0 & \cdots & -1 & 1
    \end{pmatrix}. 
  \end{align*} 
  Then, 
  \begin{align*}
    \nu 
    = \begin{pmatrix} 
      \theta''_2 - \theta''_1 - (\theta'_2 - \theta'_1) \\
      \theta''_3 - \theta''_2 - (\theta'_3 - \theta'_2) \\
      \vdots \\
      \theta''_\tau - \theta''_{\tau-1} - (\theta'_\tau - \theta'_{\tau-1})
    \end{pmatrix}
  \end{align*}
  which is an alternate definition of the vector of causal effects (comparing adjacent time points instead of comparing to baseline), that was introduced previously.
\end{example}

We can establish a weak convergence result for an estimator of $\nu$ that is analogous to Theorem~\ref{thm:weak_converg_theta}. The SDR estimator for $\nu$ is given by $\widehat{\nu}_n = K \widehat{\theta}_n \in \mathbbm{R}^{k \times 1}$. By \eqref{eqn:mvn_theta_hat}, the approximate distribution of $\widehat{\nu}_n$ is also a multivariate normal distribution. Specifically, 
\begin{align}
  \widehat{\nu}_n \overset{app}{\sim} N_{k}(\nu, S^*_n),
  \label{eqn:mvn_nu_hat}
\end{align}
where $S^*_n = K S_n K^T \in \mathbbm{R}^{k \times k}$ such that 
\begin{align}
  n S^*_n 
  &= n K S_n K^T\nonumber\\
  &= K (n S_n) K^T\nonumber\\
  &\overset{p}{\longrightarrow} K \Sigma K^T = \Sigma^* \in \mathbbm{R}^{k \times k}
  \label{eqn:S*_converg}
\end{align}
by \eqref{eqn:S_converg}. Consider the standardized quantity for $\widehat{\nu}_n$ given by
\begin{align*}
  T^*_n = (D^*_n)^{-1/2}(\widehat{\nu}_n - \nu) \in \mathbbm{R}^{k \times 1}.
\end{align*}
An estimator for the correlation matrix of $T^*_n$ is given by $R^*_n \in \mathbbm{R}^{k \times k}$, which is defined as 
\begin{align*}
  R^*_n = (D^*_n)^{-1/2} S^*_n (D^*_n)^{-1/2},
\end{align*}
where $D^*_n = \text{diag}(S^*_n) \in \mathbbm{R}^{k \times k}$ is the diagonal matrix of $S^*_n$ such that 
\begin{align}
  n D^*_n
  &= n \text{diag}(S^*_n)\nonumber\\
  &= \text{diag}(n S^*_n)\nonumber\\
  &\overset{p}{\longrightarrow} \text{diag}(\Sigma^*) = D^* \in \mathbbm{R}^{k \times k}
  \label{eqn:D*_converg}
\end{align}
by \eqref{eqn:S*_converg}. Then, by \eqref{eqn:mvn_nu_hat}, the approximate distribution of $T^*_n$ is once again a multivariate normal distribution. Specifically,
\begin{align}
  T^*_n \overset{app}{\sim} N_{k}(0, R^*_n).
  \label{eqn:mvn_T*}
\end{align}

Finally, notice that Slutsky's Theorem can be used again since
\begin{align*}
  R^*_n &= (D^*_n)^{-1/2} S^*_n (D^*_n)^{-1/2} \\
  &= (n D^*_n)^{-1/2} (n S^*_n) (n D^*_n)^{-1/2} \\
  &\overset{p}{\longrightarrow} (D^*)^{-1/2} \Sigma^* (D^*)^{-1/2} = R^* \in \mathbbm{R}^{k \times k}
\end{align*}
by \eqref{eqn:S*_converg} and \eqref{eqn:D*_converg} and that 
\begin{align*}
  T^*_n 
  &= (D^*_n)^{-1/2}(\widehat{\nu}_n - \nu) \\
  &= (n D^*_n)^{-1/2} \sqrt{n} (\widehat{\nu}_n - \nu) 
\end{align*}

\begin{theorem}[Weak convergence of SDR estimator for $\nu$]
  \label{thm:weak_converg_nu}
  Assume conditions of Theorem~\ref{thm:weak_converg_theta} hold. Then,
  \begin{align*}
    T^*_n = (n D^*_n)^{-1/2} \sqrt{n} (\widehat{\nu}_n - \nu) \overset{d}{\longrightarrow} N_{k}(0, R^*).
  \end{align*}
  
\end{theorem}

\subsection{Global Inference}

As previously described, we are interested in testing for a global difference in the rate of change of the natural outcome trajectory $\overline{\theta}'$ versus the counterfactual outcome trajectory $\overline{\theta}''$. We use Theorem~\ref{thm:weak_converg_nu} to establish a general global hypothesis testing procedure that can be used to conduct not only this particular global test but also other relevant global tests.

Consider the following global hypothesis 
\begin{align}
  &H_0: \nu = h\nonumber\\
  &H_1: \nu \neq h,
  \label{eqn:hyp_global}
\end{align}
where 
\begin{align*}
  \nu = 
  \begin{pmatrix} 
    \nu_1 & \nu_2 & \cdots & \nu_k 
  \end{pmatrix}^T \in \mathbbm{R}^{k \times 1}
\end{align*}
and
\begin{align*}
  h = 
  \begin{pmatrix} 
    h_1 & h_2 & \cdots & h_k 
  \end{pmatrix}^T \in \mathbbm{R}^{k \times 1}
\end{align*}
is a user-given vector of constants that is oftentimes set equal to the zero vector $0$.

\begin{example}[Global test for difference in rate of change]
  \label{ex:hyp_global_roc}
  Let $K$ be the matrix in Example~\ref{ex:K_roc} and $h$ be the $0$ vector. Then, the global null hypothesis becomes
  \begin{align*}
    H_0: \Delta_2 = \cdots = \Delta_\tau = 0
  \end{align*}
  Testing this global hypothesis is equivalent to testing whether there is a global difference in the rate of change of the natural outcome trajectory $\overline{\theta}'$ and counterfactual outcome trajectory $\overline{\theta}''$.
\end{example}

To establish global tests for the hypothesis in \eqref{eqn:hyp_global}, notice that under $H_0$,
\begin{align}
  T^*_n = (D^*_n)^{-1/2} (\widehat{\nu}_n - h) \overset{d}{\longrightarrow} N_{k}(0, R^*)
  \label{eqn:mvn_T*_H0}
\end{align}
based on Theorem~\ref{thm:weak_converg_nu}. Therefore, standard global hypothesis tests can be constructed. One such test that we consider is a Wald test. A Wald test follows trivially from \eqref{eqn:mvn_T*_H0} by application of Slutsky's Theorem.

\begin{corollary}[Global Wald test]
  \label{cor:global_wald}
  Suppose $R^*_n$ is a consistent estimator of the correlation matrix $R^*$. A Wald test statistic is defined as
  \begin{align*}
    T^*_w = (T^*_n)^T (R^*_n)^{-1} (T^*_n).
  \end{align*} 
  Under $H_0$ in \eqref{eqn:hyp_global}, the limiting distribution of $T^*_w$ is a chi-square distribution with $k$ degrees of freedom $\chi^2_k$, or
  \begin{align*}
    T^*_w \overset{d}{\longrightarrow} \chi^2_k.
  \end{align*}
  
  Let $t^*_w$ be the observed Wald test statistic. $H_0$ is rejected at the significance level $\alpha$ if $t^*_w > q_{w,\alpha}$, where $q_{w,\alpha}$ is the $(1-\alpha)$ quantile of $\chi^2_k$, that is, $P(\chi^2_k \leq q_{w,\alpha}) = 1-\alpha$. The global p-value is calculated as $p_w = P(\chi^2_k > t^*_w)$.
\end{corollary}

An alternative global test for the hypothesis in \eqref{eqn:hyp_global} that we also consider is a maximum test. Under $H_0$, consider 
\begin{align*}
  T^*_n = 
  \begin{pmatrix}
    T^*_{1,n} & T^*_{2,n} & \cdots & T^*_{k,n}
  \end{pmatrix},
\end{align*}
where $T^*_{j,n}$ is the $j$th component of $T^*_n$. Denote the maximum of $|T^*_{j,n}|$, $j = 1, \ldots, k$ by $\text{max}(|T^*_n|)$. Finally, notice that 
\begin{align*}
  P(\text{max}(|T^*_n|) \leq t) 
  &= P(|T^*_n| \leq t) \\
  &= P(-t \leq T^*_n \leq t).
\end{align*}
Therefore, the limiting distribution of $\text{max}(|T^*_n|)$ follows trivially from \eqref{eqn:mvn_T*_H0}.

\begin{corollary}[Global maximum test]
  \label{cor:global_max}
  A maximum test statistic is defined as 
  \begin{align*}
    T^*_m = \text{max}(|T^*_n|)
  \end{align*}
  Under $H_0$ in \eqref{eqn:hyp_global}, the limiting distribution of $T^*_m$ is given by
  \begin{align*}
    P(T^*_m \leq t) \overset{d}{\longrightarrow} \int_{-t}^t \cdots \int_{-t}^t f(x_1, \ldots, x_k; R^*_n) dx_1 \cdots dx_k \coloneqq g(t; R^*_n)
  \end{align*}
  where $f(x_1, \ldots, x_k; R^*_n)$ is the density for a multivariate normal distribution with mean $0$ and correlation matrix $R^*_n$.
  
  Let $t^*_m$ be the observed maximum test statistic. $H_0$ is rejected at the significance level $\alpha$ if $t^*_m > q_{m,\alpha}$, where $q_{m,\alpha}$ is the $(1-\alpha)$ quantile of the limiting distribution of $T^*_m$, that is, $g(q_{m,\alpha}; R^*_n) = 1-\alpha$. The global p-value is calculated as $p_m = 1 - g(t^*_m; R^*_n)$.
\end{corollary}

While we approximate the distribution of the maximum test statistic analytically, we could alternatively approximate it using a multiplier bootstrap, as in \citet{kennedy2019nonparametric}. This would allow us to avoid having to explicitly integrate a possibly high-dimensional multivariate normal distribution and thus might be computationally advantageous when the number of time points or contrasts being tested is larger. In the longitudinal settings in public health and medicine that motivate our work, the number of time points is typically modest, so we do not anticipate computational gains from using the multiplier bootstrap approximation instead of the analytic approximation.

Both the Wald and maximum tests are valid hypothesis tests to use for global inference. However, the latter test can be extended to construct simultaneous inference procedures that control the overall type I error without being overly conservative. We discuss this extension for simultaneous inference more in the next section. Although we choose to focus on a Wald test and maximum test, other global hypothesis tests can be developed within our framework.

\subsection{Simultaneous Inference}

When the global null hypothesis is rejected, it is natural to proceed with testing each individual hypothesis to see where exactly the difference lies. For example, if we find a global difference in the rate of change comparing the natural outcome trajectory $\overline{\theta}'$ with the counterfactual outcome trajectory $\overline{\theta}''$, then we would want to test for local differences. We use results from Corollary~\ref{cor:global_max} to establish a general local hypothesis testing procedure. Additionally, we extend these results to construct simultaneous confidence intervals.

Consider $k$ local hypotheses for $j = 1,\ldots, k$
\begin{align}
  &H_0^j: \nu_j = h_j\nonumber\\
  &H_1^j: \nu_j \neq h_j,
\label{eqn:hyp_local}
\end{align}
where $\nu_j$ and $h_j$ are the $j$th components of $\nu$ and $h$, respectively.

Consider $T^*_{j,n}, j = 1, \ldots, k$. Based on \eqref{eqn:mvn_T*_H0} and under $H^j_0$, a local test is given by 
\begin{align}
  T^*_{j,n} = (D^*_n)_{jj}^{-1/2} (\widehat{\nu}_{j,n} - h_j) \overset{d}{\longrightarrow} N(0,1),
  \label{eqn:local_test}
\end{align}
where $(D^*_n)_{jj}$ is the matrix component corresponding the $j$th row and $j$th column of $D^*_n$ and is equal to $(S^*_n)_{jj}$. This local test has rejection threshold $z_{\alpha/2}$, which is the $(1-\alpha/2)$ quantile of a standard normal distribution $Z$, that is, $P(Z \leq z_{\alpha/2}) = 1 - \alpha/2$. For $j = 1, \ldots, k$, it yields observed test statistic $t^*_j$ with p-value $p_j = P(|Z| > |t^*_j|)$. 

However, when these local tests are conducted simultaneously, the probability of falsely rejecting at least one true null hypothesis becomes larger than the nominal significance level $\alpha$. This is the origin of the multiple comparisons problem. To ensure that the overall type I error, also known as the family wise error rate, is bounded by $\alpha$, we can use a single-step multiple testing procedure to identify a new threshold for rejection and calculate adjusted p-values. Bounding the overall type I error by $\alpha$ means choosing a new rejection threshold $c_{\alpha}$ such that
\begin{align*}
  P\left(\left|t^*_j\right| \leq c_{\alpha} \ \text{for all} \ j \ \mid H_0\right) \geq 1 - \alpha.
\end{align*} 

We first consider a simple procedure that is frequently employed known as the Bonferroni procedure. To motivate the Bonferroni procedure, notice that the following statement
\begin{align*}
  1-\alpha = P\left(\left|t^*_j\right| \leq c_{b, \alpha} \ \text{for all} \ j \right) \geq 1 - \sum_{j = 1}^k P\left(\left|t^*_j\right| > c_{b, \alpha}\right).
\end{align*}
holds when $c_{b,\alpha} = z_{\alpha/(2k)}$. Thus, the Bonferroni-adjusted p-values are calculated by multiplying the unadjusted p-values with the number of comparisons $k$, that is, $p_{b,j} = k \times p_j, j = 1, \ldots, k$.

Since the Bonferroni procedure is free of distributional assumptions and does not assume any dependence between tests, it is generally a conservative procedure, meaning that it oftentimes yields an overall type I error that is less than $\alpha$. This motivates us to consider an alternative multiple testing procedure that is generally less conservative. We refer to this alternative procedure as the maximum procedure since it is developed using results from Corollary~\ref{cor:global_max}.
To motivate the maximum procedure, notice that 
\begin{align*}
  1-\alpha &= 
  P\left(\left|t^*_j\right| \leq c_{m, \alpha} \ \text{for all} \ j \right) \\
  &= P(\text{max}(|t^*_1|, \ldots, |t^*_k|) \leq c_{m,\alpha}) \\
  &= P(t^*_m \leq c_{m,\alpha}) \\
  &\approx g(c_{m,\alpha}; R^*_n),
\end{align*}
with the approximation in the final line holding when $c_{m,\alpha} = q_{m, \alpha}$ by definition of the limiting distribution of $T^*_m$ in Corollary~\ref{cor:global_max}. Thus, the maximum-adjusted p-values are calculated as $p_{m,j} = 1-g(|t^*_j|; R^*_n), j = 1, \ldots, k$. We summarize both proposed single-step multiple testing procedures in Corollary~\ref{cor:local}.

\begin{corollary}[Local test]
  \label{cor:local}
  For $j = 1, \ldots, k$, a local test statistic is defined as 
  \begin{align*}
    T^*_{j,n}.
  \end{align*}
  Under $H_0^j$ in \eqref{eqn:hyp_local}, the limiting distribution of $T^*_{j,n}$ is a standard normal distribution $Z$, or 
  \begin{align*}
    T^*_{j,n} \overset{d}{\longrightarrow} Z.
  \end{align*}

  Let $t^*_j, j = 1, \ldots, k$ be the observed test statistics. Using the Bonferroni procedure, $H_0^j$ is rejected at the significance level $\alpha$ if $|t^*_j| > z_{\alpha/(2k)}$, and the p-values are calculated as $p_{b,j} = k \times P(|Z| > |t^*_j|), j = 1, \ldots, k$. Using the maximum procedure, $H_0^j$ is rejected if $|t^*_j| > q_{m,\alpha}$, and the p-values are calculated as $p_{m,j} = 1-g(|t^*_j|; R^*_n), j = 1, \ldots, k$.
\end{corollary}

We can extend the results in Corollary~\ref{cor:local} to construct simultaneous confidence intervals for $\nu_j, j = 1, \ldots, k$. We first consider the construction of pointwise confidence intervals, which can be done by inverting the local test for $H^j_0$ given in \eqref{eqn:local_test}. Specifically, we invert 
\begin{align*}
  P(|T^*_{j,n}| \leq z_{\alpha/2}) = 1 - \alpha,
\end{align*}
where $T^*_{j,n} = (D^*_n)_{jj}^{-1/2} (\widehat{\nu}_{j,n} - \nu_j)$, and obtain the following pointwise confidence interval for $\nu_j$
\begin{align*}
  \bigg(\widehat{\nu}_{j,n} - z_{\alpha/2} \sqrt{(D^*_n)_{jj}}, \ \widehat{\nu}_{j,n} + z_{\alpha/2} \sqrt{(D^*_n)_{jj}} \bigg).
\end{align*}
A simple approach for constructing simultaneous confidence intervals for $\nu_j, j = 1, \ldots, k$ would be to replace the original threshold $z_{\alpha/2}$ with a larger threshold $c_{\alpha}$, that is,
\begin{align*}
  \bigg(\widehat{\nu}_{j,n} - c_{\alpha} \sqrt{(D^*_n)_{jj}}, \ \widehat{\nu}_{j,n} + c_{\alpha} \sqrt{(D^*_n)_{jj}} \bigg).
\end{align*}
Both the Bonferroni and maximum rejection thresholds given in Corollary~\ref{cor:local} would be valid choices for $c_{\alpha}$. 

In this paper, we only consider two possible single-step procedures. Each of these procedures use a common threshold value for conducting simultaneous inference. Other single-step procedures, as well as step-wise procedures, which tend to be more powerful, may also be of interest.

\section{Simulation Study}
\label{sec:sim}

\begin{figure}[!tbp]
  \centering
  \includegraphics[width=\textwidth]{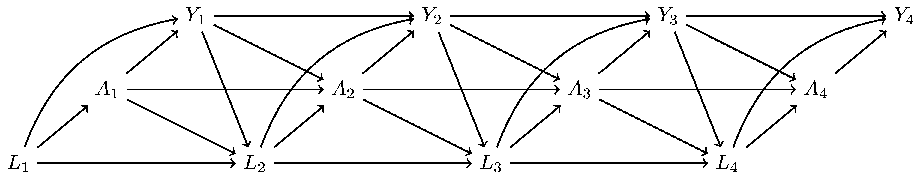}
  \caption{DAG representing data generating process when $\tau = 4$.}
  \label{fig:dag}
\end{figure}

We present a simulation study to illustrate the performance of our inference framework on balanced longitudinal data, where all individuals are assessed at the same set of pre-specified assessment times $v_t$, $t \in {1,\ldots,\tau}$. We considered a $\tau = 4$ setting with time-varying confounding and a continuous exposure. We used the directed acyclic graph (DAG) in Figure~\ref{fig:dag} to generate 1000 datasets of each sample size $n \in \{250, 500, 1000, 2500, 5000, 10000\}$ containing 3 continuous, time-varying variables: confounder $L$, exposure $A$, and outcome $Y$.

\begin{figure}[!tbp]
  \centering
  \includegraphics[width=\textwidth]{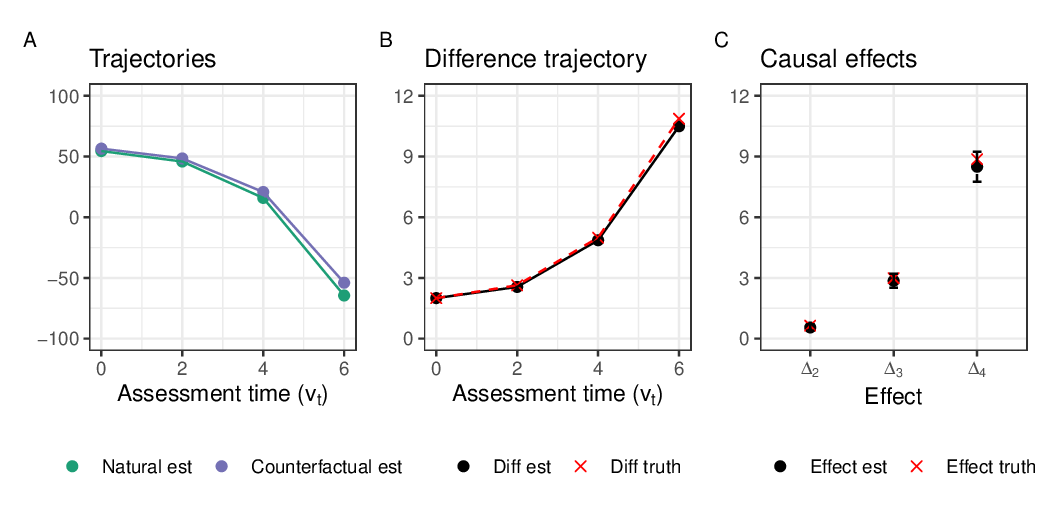}
  \caption{Results from a single simulated dataset when $\beta = 1$ and $n = 2500$. Panel A visualizes estimates of the natural outcome trajectory $\overline{\theta}'$ and counterfactual outcome trajectory $\overline{\theta}''$. Panel B visualizes an estimate of the difference trajectory and the true difference trajectory. Panel C visualizes an estimate of the vector of causal effects $\Delta$ and its simultaneous 95\% confidence intervals, as well as the true vector of causal effects.}
  \label{fig:theta_nu}
\end{figure}

We used a data generating mechanism that yields non-linear, decreasing counterfactual and natural outcome trajectories. These trajectories decrease gradually at first and then begin to decrease more rapidly, as visualized in Figure~\ref{fig:theta_nu}, reflecting a trend commonly seen in health outcomes, such as cognitive decline. For illustration purposes, we considered a simple LMTP, in which the exposure input is shifted down by 1 unit at each $t$ for all individuals, that is, $d(a_t, h_t) = a_t - 1$. The data generating mechanism is as follows 
\begin{align*}
    \begin{aligned}
        L_1 
        &\sim \texttt{Normal}\{1, 1\} \\
        A_1 \mid L_1 
        &\sim \texttt{Normal}\{8.5 - L_1, 1\} \\
        Y_1 \mid (L_1, A_1) 
        &\sim \texttt{Normal}\{70.5 + \gamma_t(-L_1 + \alpha A_1), 1\} \\
        L_t \mid (L_{t-1}, A_{t-1}, Y_{t-1}) 
        &\sim 
        \begin{aligned}[t]
            \texttt{Normal}\{&
            5 + 0.47L_{t-1} - 0.24A_{t-1} \\
            &- 0.05Y_{t-1} - 0.3v_t,\ 1\}
            \ \text{for } t \in \{2,3,4\}
        \end{aligned} \\
        A_t \mid (L_t, A_{t-1}, Y_{t-1}) 
        &\sim 
        \begin{aligned}[t]
            \texttt{Normal}\{&
            10 - 0.2L_t + 0.1A_{t-1} - 0.05Y_{t-1} \\
            &+ 0.5v_t,\ 1\}
            \ \text{for } t \in \{2,3,4\}
        \end{aligned} \\
        Y_t \mid (L_t, A_t, Y_{t-1}) 
        &\sim 
        \begin{aligned}[t]
            \texttt{Normal}\{&
            78 + \gamma_t(-0.5L_t + \alpha A_t - 0.15Y_{t-1}) \\
            &- 0.3v_t - 0.2v_t^2 - 0.1v_t^3 \\
            &- \beta(0.1A_tv_t + 0.04A_tv_t^2 \\
            &\qquad + 0.02A_tv_t^3),\ 1\}
            \ \text{for } t \in \{2,3,4\},
        \end{aligned}
    \end{aligned}   
\end{align*}
where $\texttt{Normal}\{\mu, \sigma^2 \}$ is a normal distribution with mean $\mu$ and variance $\sigma^2$. Here, $\gamma_t$ parametrizes outcome $Y$ so that the vector of causal effects $\Delta$ is exactly 0 when $\beta = 0$, creating a setting where the natural outcome trajectory $\overline{\theta}'$ and the counterfactual outcome trajectory $\overline{\theta}''$ are parallel and thus do not differ in terms of the rate of change. We calculate $\gamma_t$ for each $t \in \{1, \ldots, \tau\}$ recursively, starting with $\gamma_1$. More specifically, we analytically calculate the expected natural outcome $\theta_1'$ and the expected counterfactual outcome $\theta_1''$ with $\gamma_1$ temporarily set to 1, calculate $\gamma_1 = \frac{-\alpha}{\theta_1'' - \theta_1'}$, update $\theta_1'$ and $\theta_1''$ based on this new $\gamma_1$, and repeat this entire process for the following $t \in \{2, \ldots, \tau\}$. Therefore, $\alpha$ represents the difference between $\theta_t'$ and $\theta_t''$ at each $t$ in the setting where $\overline{\theta}'$ and $\overline{\theta}''$ are parallel and there is no difference in the rate of change. Additionally, $\beta$ captures the magnitude of interaction between assessment time $v$ and exposure $A$ and indirectly controls how different $\overline{\theta}'$ is from $\overline{\theta}''$ in terms of the rate of change. When $\beta = 0$, the null hypothesis $H_0$ in \eqref{eqn:hyp_global} holds, as $\overline{\theta}'$ and $\overline{\theta}''$ are parallel. When $\beta > 0$, the alternative hypothesis $H_1$ holds, as $\overline{\theta}'$ and $\overline{\theta}''$ are non-parallel, with higher values yielding greater differences in the rates of change. Therefore, $\beta$ can be thought of as an arbitrary ``effect size'' for the difference in the rate of change. Since all variables are generated using normal distributions and are associated linearly with each other, analytical calculation of the true $\overline{\theta}'$ and $\overline{\theta}''$ follows straightforwardly by application of linearity of expectation. For our simulation, we let $\alpha = -2$ and $v_t \in \{0, 2, 4, 6\}$.

\begin{figure}[!htbp]
  \centering
  \includegraphics[width=\textwidth]{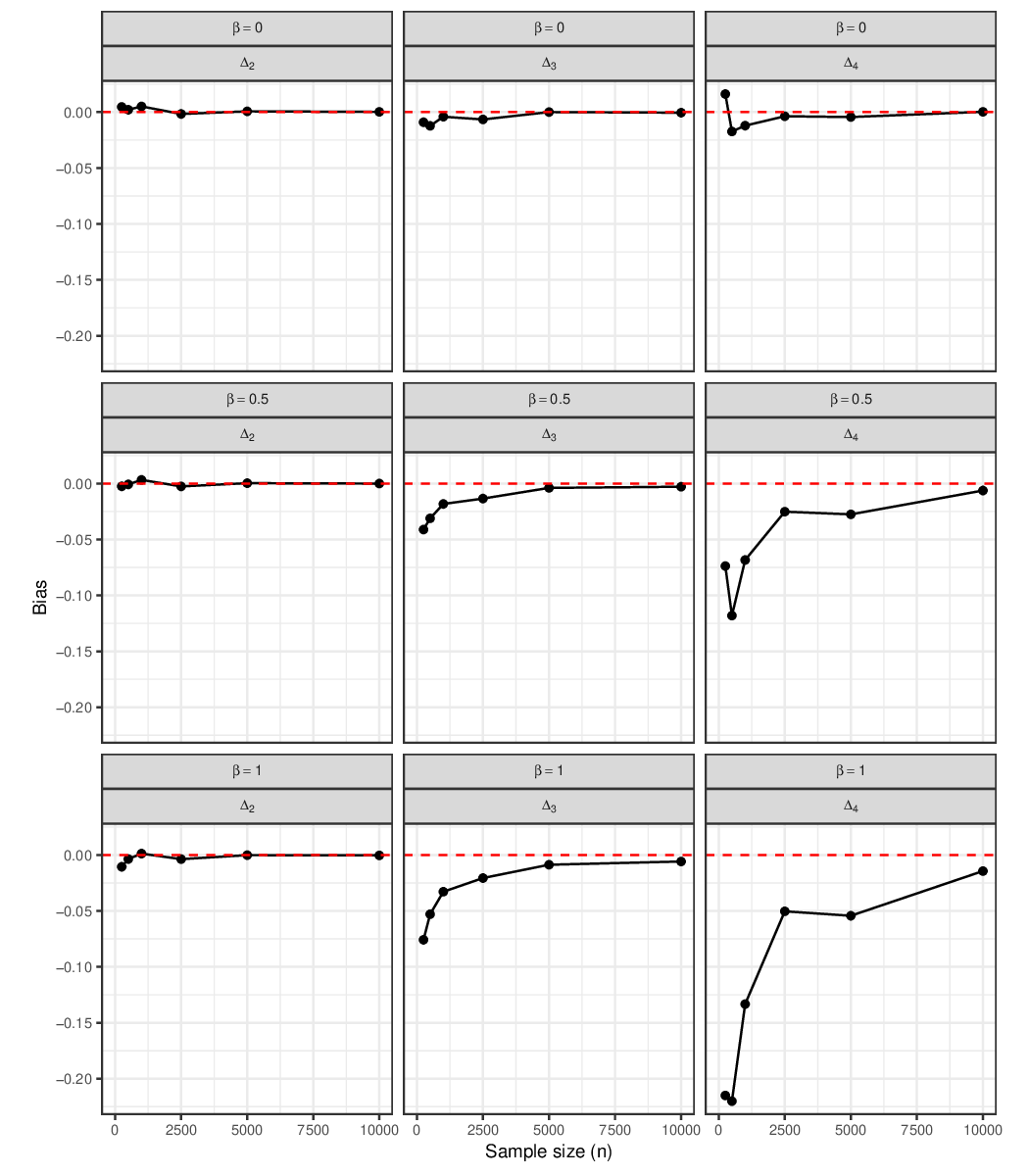}
  \caption{Bias of the SDR estimator for the vector of causal effects $\Delta$ across 1000 simulated datasets for a sequence of $n$ and $\beta \in \{0, 0.5, 1\}$.}
  \label{fig:bias}
\end{figure}

We assessed the performance of our inference framework for investigating whether the natural outcome trajectory $\overline{\theta}'$ differs from the counterfactual outcome trajectory $\overline{\theta}''$ in terms of the rate of change. For each $n$, we repeated our simulation for a sequence of $\beta \in [0, 1]$ to assess performance under both the null and alternative hypotheses. Figure~\ref{fig:theta_nu} presents estimates of the natural outcome trajectory $\overline{\theta}'$ and the counterfactual outcome trajectory $\overline{\theta}''$, as well as the corresponding vector of causal effects $\Delta$, from a single simulated dataset when $\beta = 1$ and $n = 2500$. The simultaneous 95\% confidence intervals of $\Delta$ capture the true non-zero causal effects, while the global maximum test yields a significant result (test statistic of 27.5 with p-value $< 0.001$), showing that our approach for estimation and inference is performing as expected. Figure~\ref{fig:bias} presents the bias in the SDR estimator for $\Delta$ across several sample sizes $n$ when $\beta \in \{0, 0.5, 1\}$. As $n$ increases, bias converges to 0 as expected for each component of the SDR estimator of $\Delta$. Convergence is more rapid when $\beta$ is low and for components of the SDR estimator of $\Delta$ that are associated with earlier time points, suggesting that a larger sample size is needed for estimation using the Super Learner when there are stronger interaction effects and thus more variation in $Y_t$ in the underlying data generating process and when there are more time points to recurse through in the sequential regression algorithm used for estimation. 

\begin{figure}[!htbp]
  \centering
  \includegraphics[width=\textwidth]{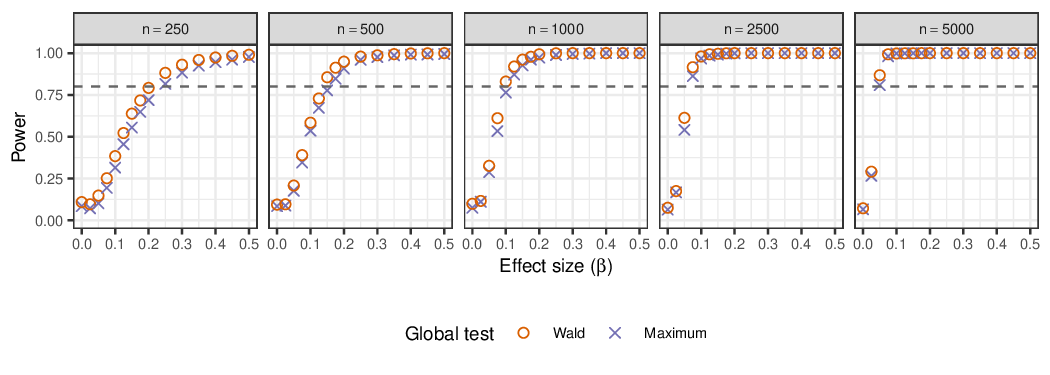}
  \caption{Power of the global Wald and maximum tests across 1000 simulated datasets for a sequence of $\beta$ and several $n$. The dashed, black line represents the threshold for a power of 0.80.}
  \label{fig:global_power}
\end{figure}

\begin{figure}[!htbp]
  \centering
  \includegraphics[width=\textwidth]{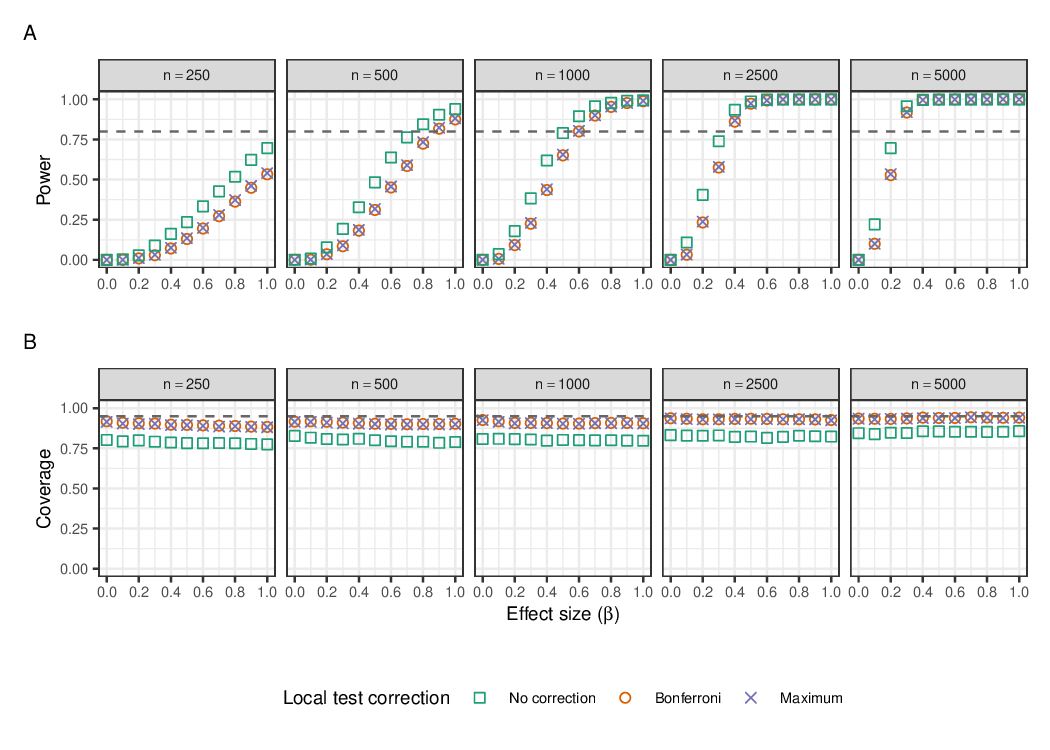}
  \caption{Simultaneous power of the local tests (Panel A) and simultaneous coverage of the confidence intervals of the vector of causal effects $\Delta$ (Panel B) under no multiple testing correction and under the Bonferroni and maximum procedures across 1000 simulated datasets for a sequence of $\beta$ and several $n$. The dashed, black line represents the threshold for a power of 0.80 (Panel A) or a coverage of 0.95 (Panel B).}
  \label{fig:simult_comb}
\end{figure}

Figure~\ref{fig:global_power} presents the power of the global Wald and maximum tests across a sequence of $\beta$ and for several $n$. Here, power is defined as the probability of rejecting the null hypothesis $H_0$. The result at $\beta = 0$ is interpreted as the type I error since it indicates the probability that the null hypothesis is falsely rejected. Figure~\ref{fig:simult_comb} presents results under no multiple testing correction and under the Bonferroni and maximum procedures for the simultaneous power of the local tests and the simultaneous coverage of the confidence intervals of the vector of causal effects $\Delta$. Here, simultaneous power is defined as the probability that $H_0^j$ is rejected for all $j = 2, \ldots, \tau$, with the result at $\beta = 0$ interpreted as the simultaneous type I error. Simultaneous coverage is defined as the probability that the confidence interval for $\Delta_j$ covers the true value for all $j = 2, \ldots, \tau$. Subtracting the simultaneous coverage result at $\beta = 0$ from 1 yields the overall type I error, or family wise error rate, as it corresponds to the probability of falsely rejecting $H_0^j$ for at least one local test. The results in Figures~\ref{fig:global_power} and~\ref{fig:simult_comb} highlight several important characteristics. First, as $\beta$ and $n$ increase, power increases as expected for both the global and local tests. Second, the global Wald test tends to be slightly more powerful than the global maximum test. Third, when the data generating mechanism yields local tests that are poorly correlated, such as in this simulation, then the maximum procedure displays no advantage over the Bonferroni procedure in terms of simultaneous power and coverage. Fourth, using the empirical covariance matrix of the estimated EIFs to conduct inference yields anti-conservative confidence intervals and global tests with slightly below nominal simultaneous coverage and above nominal type I error. Previous work has shown similar results in a longitudinal setting when using the empirical variance of the EIF as an estimator for the variance of doubly robust estimators like the SDR estimator and has highlighted that this behavior is exacerbated as positivity violations increase \citep{tran2023robust}. This is because the covariance matrix and the variance are themselves estimands that have a non-negligible first order bias term, but plug-in estimators, such as the empirical covariance matrix, do no correct for this bias term.

\section{Illustrative Application: The Effects of Shifted Blood Pressure on Progression of Dementia}
\label{sec:app}

We apply our proposed inference framework to investigate whether a longitudinal shift intervention on blood pressure affects the progression of dementia using the Columbia Brain Health DataBank (CBDB). The CBDB is an integrated data resource that combines Alzheimer’s disease (AD) research cohort data, electronic health records (EHR), neuroimaging, and biomarker information across the life course at Columbia University Irving Medical Center to understand the progression of  Alzheimer’s disease and related dementias.  1,084 of the individuals in the CBDB are participants of the Columbia Alzheimer’s Disease Research Center (ADRC). Clinicians and research staff in the ADRC have conducted annual standardized cognitive evaluations on individuals with different etiologies of dementia, as well as those without a dementia diagnosis, since 2005 \citep{beekly2007national}. Our analysis focuses on 995 participants from this EHR-linked ADRC cohort.

Although cognitive evaluations are intended to be measured annually, the data is subject to complex patterns of missingness due to loss-to-follow up, drop-out/re-entry, refusal to respond, etc. As a result, there are steep drop offs in the number of participants available for analysis at each annual visit following the baseline visit. For example, from the 995 individuals with baseline visit 0, only 601 have visit 1, 513 have visit 2, and so on. In order to construct a sample over a 6-year time period that is sufficiently large enough for LMTP estimation, we binned the annual visits into 3 time-bins: baseline, 1-3 years since baseline, and 4-6 years since baseline. In cases with multiple visits in a bin, the latest visit was used. Here, we are defining the causal estimand $\Delta_t$ with respect to the binned visit indices $t \in \{1, 2, 3\}$ and effectively assuming that variation within these observation bins is unimportant for our purposes. The coarsening of observations into these bins is admittedly somewhat arbitrary and is not based on clinical considerations, so the results that we generate in this section should be interpreted only as illustrative of the proposed methodology. 

Additionally, since the current LMTP methodology only accommodates missingness from loss-to-follow up, we excluded all participants with any other patterns of missingness, leaving a total of 818 individuals in the final analysis. Extending our set up to accommodate loss-to-follow up follows straightforwardly from results established in \citet{diaz2024causal}. We now define $\theta_t$ as the expected counterfactual outcome at binned visit $t$ under an intervention defined by function $d$ in a hypothetical world where there is no loss to follow-up. As a result, we must multiply the exposure density ratio $r_{s,t}$ by an indicator for being lost to follow-up and the inverse probability of being lost to follow-up. We denote this new weight by $w_{s,t}$ for $s \in \{1,\ldots,t\}$. The \texttt{lmtp} R package accommodates loss-to-follow up by allowing users to specify indicator variables, where 0 denotes being lost to follow-up at the next binned visit. 

In our analysis, the time-varying exposure of interest is systolic blood pressure, and the outcome of interest is the Clinical Dementia Rating (CDR) Sum of Boxes score. Clinicians administer the CDR to patients to evaluate 6 cognitive domains (i.e., memory, orientation, judgment and problem solving, community affairs, home and hobbies, and personal care) \citep{hughes1982new}. The CDR yields both a global and Sum of Boxes score. The Sum of Boxes score ranges from 0 to 18 (with higher scores indicating worse performance) and is calculated by summing up the patient's scores across the 6 domains. Studies have found that the Sum of Boxes score provides a finer gradation of impairment than the global score and is a promising primary endpoint for AD trials \citep{lynch2005clinical, coley2011suitability}. Studies have also demonstrated that the Sum of Boxes score can be used to stage dementia severity \citep{o2008staging, o2010validation}. Finally, several studies have examined the rate of change in the Sum of Boxes score as a measure of the progression of dementia or AD \citep{rossetti2010cerad, tschanz2011progression, williams2013progression}. High systolic blood pressure in participants with a diagnosis of mild cognitive impairment (MCI) was associated with faster increases in the Sum of Boxes score (or in other words, faster progression of dementia) in the National Alzheimer’s
Coordinating Center (NACC) Uniform Data Set (UDS) \citep{goldstein2013high}. 

We view both blood pressure and cognitive function as processes that unfold over time. Specifically, we view blood pressure measured at binned visit $t$ as reflecting vascular status over the preceding months, and we assume this state precedes and may influence the Sum of Boxes score recorded at that visit. We selected a small set of baseline and time-varying confounders for the analysis based on our subject-matter knowledge. The time-varying confounders include body mass index (BMI) and time since baseline, while the baseline confounders include age at baseline, gender (categorized as male or female), and race/ethnicity (categorized as white non-Hispanic, Black non-Hispanic, Hispanic/Latino of any race, or other). A descriptive summary of the exposure, outcome, and covariates in our study sample is provided in the \hyperref[sec:supplement]{Supporting Information}.

When a patient's blood pressure is high, there are several possible interventions a physician may consider, including recommending lifestyle modifications and prescribing medications. We evaluate an additive shift LMTP that reduces the systolic blood pressure input by shift $\delta$ only if it is greater than or equal to threshold $\gamma$, or 
\begin{align*}
    d(a_t, h_t)= \begin{cases} a_t - \delta, & \text { if } a_t \geq \gamma \\ a_t, & \text { if } a_t < \gamma \end{cases}.
  \end{align*}

Since physicians often recommend interventions when systolic blood pressure exceeds 130, the threshold for stage 1 hypertension, we focus primarily on the scenario with $\delta = 10$ and $\gamma = 130$. Here, $\gamma$ is chosen so that only individuals with sufficiently large systolic blood pressure are subject to downwards shifts, and $\delta$ is chosen to be small relative to the empirical range of systolic blood pressure. This ensures that the shifted values $a_t - \delta$ remain within the observed support, mitigating potential positivity concerns by design. Additionally, to assess positivity visually, we present histograms in the \hyperref[sec:supplement]{Supporting Information} of the cumulative estimated weights, defined for each $\theta_t$ under the LMTP as $W_{s,t} = \prod_{k=1}^s w_{k,t}$ from $s \in \{1,\ldots,t\}$. To avoid confusion, we have excluded zero weights arising by construction due to the censoring indicator. These histograms show no extreme weights, providing no clear evidence of practical positivity violations.

\begin{figure}[!htbp]
  \centering
  \includegraphics[width=\textwidth]{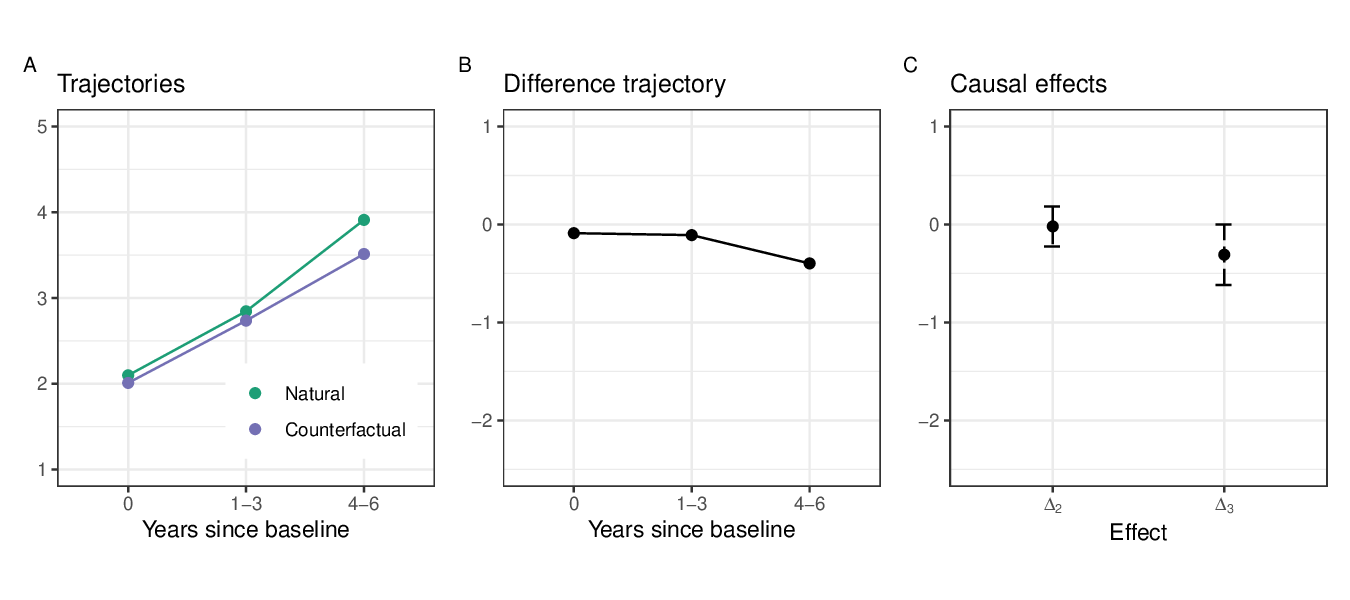}
  \caption{Results from the illustrative application examining the effect of shifting blood pressure on the progression of dementia. The LMTP of interest reduces the systolic blood pressure input by shift $\delta = 10$ only if it is greater than or equal to threshold $\gamma = 130$. Panel A visualizes estimates of the natural outcome trajectory $\overline{\theta}'$ and counterfactual outcome trajectory $\overline{\theta}''$. Panel B visualizes an estimate of the difference trajectory. Panel C visualizes an estimate of the vector of causal effects $\Delta$ and its simultaneous 95\% confidence intervals.}
  \label{fig:applied_del10_gam130}
\end{figure}

Figure~\ref{fig:applied_del10_gam130} presents results evaluating whether the LMTP with $\delta = 10$ and $\gamma = 130$ affects the rate of change in the Sum of Boxes score, as compared with no intervention. Visual inspection suggests that the estimated counterfactual outcome trajectory increases less steeply than the natural outcome trajectory. Since the global maximum test yields a test statistic of 2.220 ($\text{p-value} = 0.0497$), we conclude that the two trajectories differ significantly in terms of their rates of change at the 0.05 significance level. Based on the maximum procedure for multiple testing, only the causal effect comparing 4-6 years later to baseline is significantly different from 0 ($\text{p-value} = 0.0497$). More specifically, the increase in the Sum of Boxes score from baseline to 4-6 years later is 0.309 (95\% CI: 0.0003, 0.617) points lower, on average, under the LMTP versus no intervention. These results suggest that a shift intervention reducing systolic blood pressure may be protective against the progression of dementia, with benefits only appearing if enough time has passed.

Since one could envision a more aggressive intervention (i.e., with a greater shift or lower threshold), results for $\delta = 20$ and $\gamma = 130$, as well as for $\delta = 10$ and $\gamma = 120$, are also provided in the Supplementary Materials to demonstrate the impact of modifying the chosen LMTP on our conclusions. Decreasing the threshold to $\gamma = 120$ has little impact on the estimated causal effects but yields a noticeable increase in their standard errors such that the global maximum test is no longer significant at the 0.05 significance level (test statistic of 1.669 with $\text{p-value} = 0.160$). Increasing the shift to $\delta = 20$ leads to more negative estimates of the causal effects, especially comparing 4-6 years later to baseline, reflecting greater divergences between the natural and counterfactual outcome trajectories. It also yields even greater increases in the standard errors of the estimated causal effects. Once again, the global maximum test is no longer significant at the 0.05 significance level (test statistic of 2.070 with $\text{p-value} = 0.067$). These results highlight that even small modifications to the shift intervention under consideration may alter conclusions, particularly when the sample size is low.

\section{Discussion}
\label{sec:discussion}

When analyzing longitudinal data, scientific interest may lie in investigating the effect of an intervention on an outcome's rate of change. In a health context, this could entail studying the progression (or trajectory) of a disease over time under different hypothetical interventions. Here, we bridge the gap in the application of causal inference methodology to these longitudinal scientific questions. Specifically, we extend the LMTP methodology to target effects of complex, exposure-dependent interventions on rates of change in an outcome over time. We also propose a novel inference framework to formally test hypotheses about whether these interventions affect the outcome's trajectory. 

By building off the LMTP methodology, our approach is flexible and inherits many desirable properties, as already discussed.  Our inference framework exploits the theoretical guarantees of the EIF-based SDR estimator to conduct global and local hypothesis tests and construct simultaneous confidence intervals for linear functions of the stacked vector of counterfactual outcome trajectories $\theta$. Flexibility in the choice of the $K$ matrix to define the linear function enables our framework to be used for a variety of causal quantities and hypothesis tests of interest. We have illustrated the utility of our framework in investigating whether a longitudinal shift intervention changes the counterfactual outcome trajectory, as compared with no intervention. 

Our approach has some limitations, and questions remain open for future work. First, the LMTP estimation procedure is computationally intensive. This is in part due to the use of the Super Learner for nuisance parameter estimation and could be mitigated by using a more computationally efficient data-adaptive technique. However, more fundamentally, it is due to the use of an estimation algorithm that recurses backwards through time and must be run $\tau$ times to estimate the entire counterfactual outcome trajectory under a specific intervention. \citet{susmann2025computationally} recently proposed an approach for estimating the entire counterfactual outcome trajectory using a single algorithmic run to reduce computation time. Their approach relies on a time-smoothed SDR estimator that pools sets of sequential regressions to reduce the total number of regressions required. Our inference framework is based on a joint weak convergence result for $\theta$ and can thus accommodate any alternative estimator that admits analogous conditions, including a joint weak convergence result and a consistent covariance estimator (e.g., the time-smoothed SDR estimator). In principle, we could swap the SDR estimator in our framework with the time-smoothed SDR estimator without changing any of the remaining theory. However, even with such modifications, substantial computational challenges will persist when the number of time points $\tau$ is large. We recommend our approach primarily for longitudinal studies commonly encountered in public health and medicine, where $\tau$ is typically modest (e.g., $\leq 10$), rather than for settings with large $\tau$ (e.g., time series data). In such settings, additional statistical challenges may also arise, as the dimensionality of the exposure and covariate histories increases and achieving the required convergence rates may depend on the feasibility of accurately estimating high-dimensional nuisance parameters.

Additionally, variance estimation of EIF-based doubly robust estimators is conventionally achieved by taking the empirical variance of the EIF. Following this practice, we use the empirical covariance matrix of the EIFs as a plug-in estimator of the covariance matrix of the SDR estimator for $\theta$ in our inference framework. However, this plug-in estimator can yield anti-conservative confidence interval coverage and type I error, as shown in our simulation. \citet{tran2023robust} demonstrated similar results when using the empirical variance of the EIF in the context of longitudinal dynamic treatment regimes and proposed two novel approaches for variance estimation. Future work could consider adapting one of these approaches to achieve more robust estimation of the covariance matrix in our inference framework. 

Finally, the current LMTP methodology does not easily accommodate certain data complications that are commonly seen in longitudinal studies. One such complication is non-monotone missingness, which can occur when there are complex patterns of drop-out and re-entry. \citet{diaz2024causal} proposed an extension that accommodates monotone loss-to-follow-up, where missingness at time $s$ implies missingness at all $t > s$. \citet{susmann2025computationally} later proposed an extension that accommodates non-monotone missingness in the outcome. However, longitudinal studies often exhibit non-monotone missingness not only in the outcome but also in the exposure and covariates (e.g., when individuals miss some clinical visits and thus contribute no data at those visits), highlighting the need for approaches that can accommodate more general missingness patterns. Another complication arises when assessment times vary among individuals rather than occurring by design at a common set of pre-specified times. The LMTP methodology requires coarsening the time scale into discrete intervals, which effectively ignores variation within intervals, even when that variation may be informative, as in our illustrative application. In a follow-up paper, we will propose an extension that better accommodates these irregular assessment times by defining estimands that take into account informative variation in observation schedules.

\section*{Author Contributions}

Anja Shahu and Daniel Malinsky conceived the project and developed statistical theory. Weijie Xia and Ying Wei provided data and guidance for the illustrative application. Anja Shahu conducted the simulation study, analyzed data for the illustrative application, and wrote the paper. Daniel Malinsky and Ying Wei reviewed the paper.

\section*{Acknowledgments}
The authors acknowledge funding from the National Institute on Aging (NIA) to support the Columbia ADRC (P30AG066462) and ADRC-EHR linkage and data pre-processing (R01AG087496), as well as from the National Center for Advancing Translational Sciences (NCATS) to support EHR data access (UL1TR001873). Anja Shahu was partially supported by T32MH135856 from the National Institute of Mental Health (NIMH). Daniel Malinsky was partially supported by K25ES034064 from the National Institute of Environmental Health Sciences (NIEHS).

\section*{Funding}
This work was supported by National Institute on Aging, P30AG066462, R01AG087496. National Center for Advancing Translational Sciences, \\ UL1TR001873, National Institute of Mental Health, T32MH135856, National Institute of Environmental Health Sciences, K25ES034064.

\section*{Disclosure}

The authors have nothing to report.

\section*{Conflicts of Interest}

The authors declare no conflicts of interests.

\section*{Data Availability Statement}
Research data are not shared.

\newpage 

\bibliographystyle{unsrtnat}
\bibliography{bibliography}

\clearpage


\setcounter{section}{0}
\setcounter{subsection}{0}
\setcounter{table}{0}
\setcounter{figure}{0}
\setcounter{equation}{0}

\renewcommand{\thesection}{A\arabic{section}}
\renewcommand{\thesubsection}{A\arabic{section}.\arabic{subsection}}
\renewcommand{\thetable}{A\arabic{table}}
\renewcommand{\thefigure}{A\arabic{figure}}
\renewcommand{\theequation}{A\arabic{equation}}

\phantomsection
\label{sec:supplement}

\title{\textit{Supplementary Materials:} Estimating Effects of Longitudinal Modified Treatment Policies (LMTPs) on Rates of Change in Health Outcomes}

\makeatletter
\let\@thanks\@empty
\makeatother

\setlength{\droptitle}{-5em}
\maketitle

\subfile{supplement}

\end{document}

%% file: supplement.tex
\begin{table}[!htbp] 
    \centering 
    \caption{Descriptive summary of the exposure, outcome, and covariates in our study sample for the illustrative application.} 
    \label{tab:adrc_summary} 
    \begin{tabular}{lccc} 
        \toprule & \makecell[c]{Visit 1\ \ \\(0 years)\ \ \\N=818} & \makecell[c]{Visit 2\ \ \\(1-3 years)\ \ \\N=567} & \makecell[c]{Visit 3\ \ \\(4-6 years)\ \ \\N=316}\\ \midrule Time of visit (years) & 0 (0) & 2 (1) & 5 (1)\\ BMI (kg/m$^2$) & 27.1 (5.0) & 27.0 (5.0) & 26.9 (5.1)\\ Age at baseline (years) & 71 (9) & 71 (8) & 71 (8)\\ Gender & & & \\ \hspace{1em}Male & 336 (41.1\%) & 226 (39.9\%) & 109 (34.5\%)\\ \hspace{1em}Female & 482 (58.9\%) & 341 (60.1\%) & 207 (65.5\%)\\ Race/ethnicity & & & \\ \hspace{1em}White non-Hispanic & 298 (36.4\%) & 218 (38.4\%) & 121 (38.3\%)\\ \hspace{1em}Black non-Hispanic & 43 (5.3\%) & 27 (4.8\%) & 13 (4.1\%)\\ \hspace{1em}Hispanic/Latino of any race & 157 (19.2\%) & 93 (16.4\%) & 49 (15.5\%)\\ \hspace{1em}Other & 320 (39.1\%) & 229 (40.4\%) & 133 (42.1\%)\\ Systolic blood pressure (mmHg) & 133.1 (18.9) & 132.3 (18.6) & 132.2 (19.1)\\ CDR Sum of Boxes score & 2.1 (3.3) & 2.8 (4.4) & 3.4 (5.6)\\ \bottomrule \multicolumn{4}{l}{\rule{0pt}{1em}\textsuperscript{1} Mean (SD); n (\%)}\\ 
    \end{tabular} 
\end{table}

\begin{figure}[!htbp]
  \centering
  \includegraphics[width=\columnwidth]{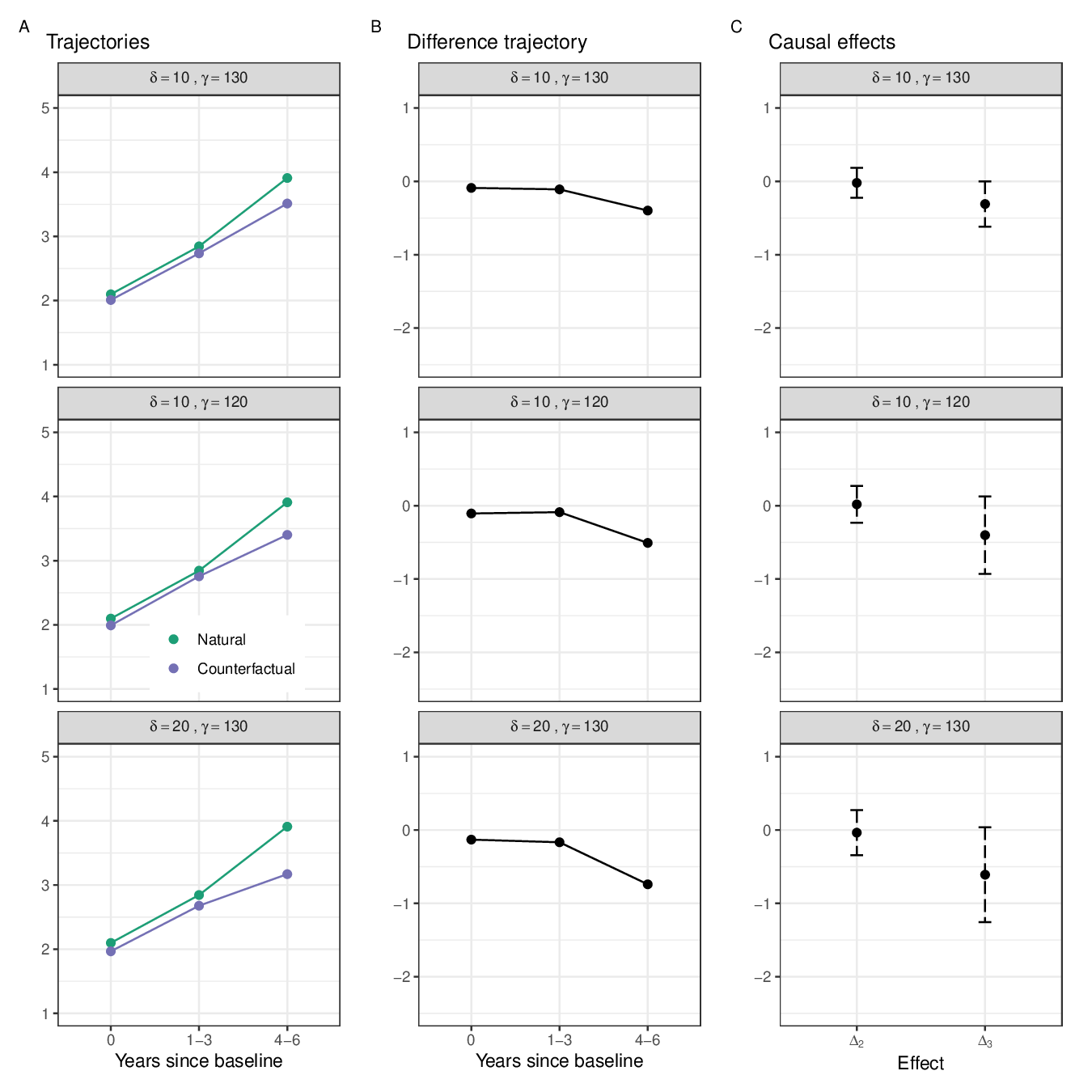}
  \caption{Results from the illustrative application examining the effect of shifting blood pressure on the progression of dementia. The LMTP of interest reduces the systolic blood pressure input by shift $\delta$ only if it is greater than or equal to threshold $\gamma$ for the following ($\delta$, $\gamma$) pairs: $\delta = 10$, $\gamma = 130$ (first row); $\delta = 10$, $\gamma = 120$ (second row); and $\delta = 20$, $\gamma = 130$ (third row). Panel A visualizes estimates of the natural outcome trajectory $\overline{\theta}'$ and counterfactual outcome trajectory $\overline{\theta}''$. Panel B visualizes an estimate of the difference trajectory. Panel C visualizes an estimate of the vector of causal effects $\Delta$ and its simultaneous 95\% confidence intervals.}
  \label{fig:applied_res_appendix}
\end{figure}

\begin{figure}[!htbp]
  \centering
  \includegraphics[width=\columnwidth]{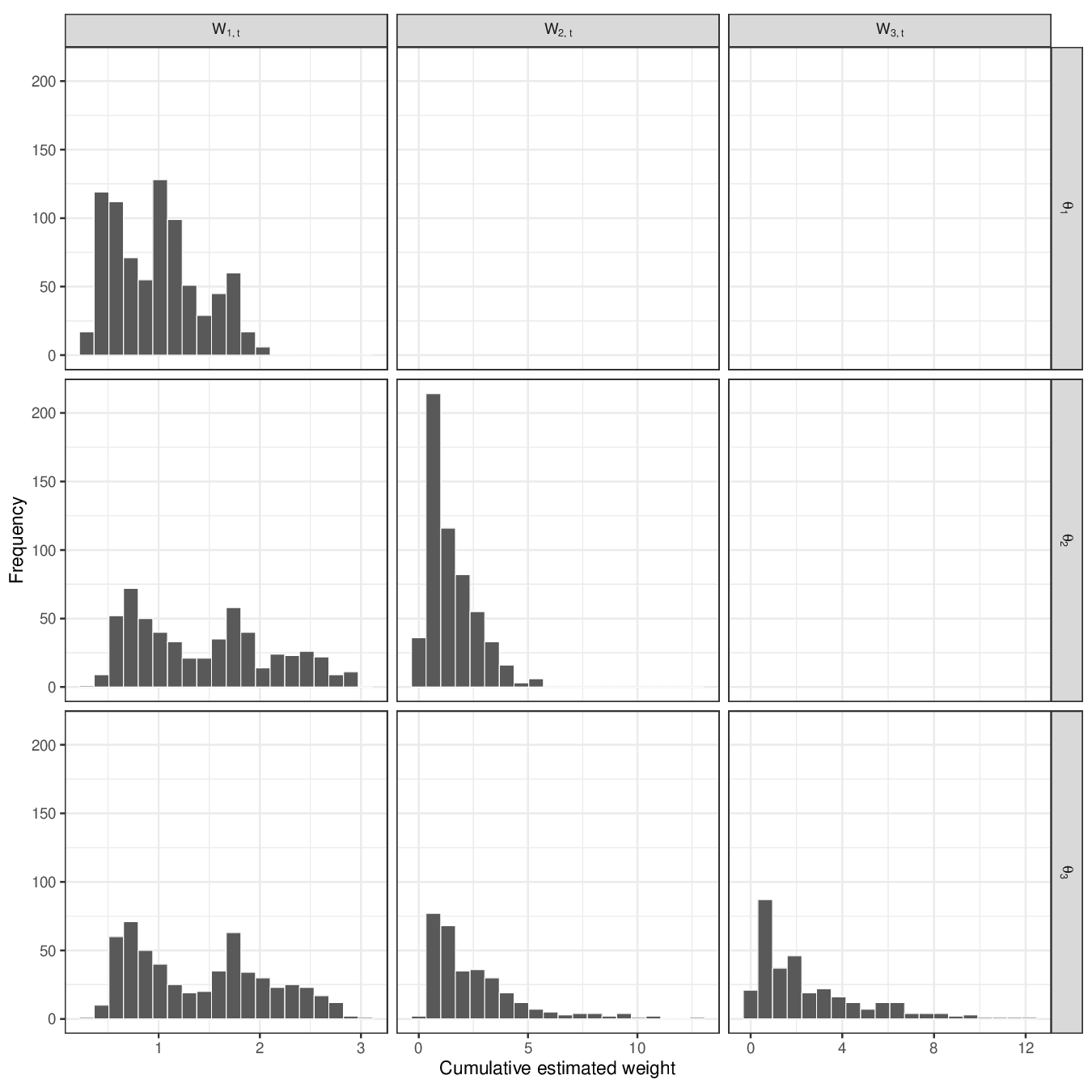}
  \caption{Cumulative estimated weights $W_{s,t}, s \in \{1,\ldots,t\}$ for each $\theta_t$ under the LMTP in the illustrative application. The LMTP of interest reduces the systolic blood pressure input by shift $\delta = 10$ only if it is greater than or equal to threshold $\gamma = 130$. Zero weights arising by construction due to the censoring indicator are excluded.}
  \label{fig:applied_cum_rat_appendix}
\end{figure}